\documentclass[12pt,english,showpacs,preprint,eqsecnum,amssymb]{revtex4}
\usepackage[T1]{fontenc}
\usepackage[latin9]{inputenc}
\usepackage{color}
\usepackage{babel}

\usepackage{amsmath}
\usepackage{graphicx}
\usepackage{amssymb}
\usepackage{esint}
\usepackage[unicode=true,pdfusetitle,
 bookmarks=true,bookmarksnumbered=false,bookmarksopen=false,
 breaklinks=false,pdfborder={0 0 1},backref=false,colorlinks=true]
 {hyperref}

\makeatletter

\providecommand{\tabularnewline}{\\}

\@ifundefined{textcolor}{}
{%
 \definecolor{BLACK}{gray}{0}
 \definecolor{WHITE}{gray}{1}
 \definecolor{RED}{rgb}{1,0,0}
 \definecolor{GREEN}{rgb}{0,1,0}
 \definecolor{BLUE}{rgb}{0,0,1}
 \definecolor{CYAN}{cmyk}{1,0,0,0}
 \definecolor{MAGENTA}{cmyk}{0,1,0,0}
 \definecolor{YELLOW}{cmyk}{0,0,1,0}
 }

\usepackage{hyperref}
\usepackage{braket}
\usepackage{setspace}

\makeatother
\def\beq{\begin{equation}}
\def\eeq{\end{equation}}
\def\beqnn{\begin{displaymath}}
\def\eeqnn{\end{displaymath}}
\def\bea{\begin{eqnarray}}
\def\eea{\end{eqnarray}}
\def\nnu{\nonumber}
\def\eno#1{Eq.~(\ref{#1})}

\def\etwo#1#2{Eqs.~(\ref{#1})\ and (\ref{#2})}

\def\fno#1{Fig.~\ref{#1}}
\def\Sno#1{Sec.~\ref{#1}}

\def\tst{\textstyle}
\def\dst{\displaystyle}

\def\baz{{\bar z}}
\def\al{\alpha}
\def\be{\beta}
\def\gam{\gamma}
\def\dta{\delta}
\def\eps{\epsilon}
\def\veps{\varepsilon}

\def\lam{\lambda}
\def\om{\omega}
\def\Gam{\Gamma}
\def\Dta{\Delta}
\def\Tta{\Theta}

\def\Lam{\Lambda}
\def\Om{\Omega}
\def\bphi{{\bar\phi}}

\def\apx{\approx}

\def\sech{{\rm sech}}

\def\ptl{\partial}

\def\part#1#2{\frac{\ptl#1}{\ptl#2}}

\def\inmipi{\int_{-\infty}^{\infty}}

\def\gtwid{\mathrel{\raise.3ex\hbox{$>$\kern-.78em\lower1ex\hbox{$\sim$}}}}
\def\ltwid{\mathrel{\raise.3ex\hbox{$<$\kern-.78em\lower1ex\hbox{$\sim$}}}}

\def\hf{\frac{1}{2}}

\def\tshf{{\tst\hf}}

\def\tily{{\tilde Y}}

\def\typ0{\tily_{\ell+1,m}}

\def\tym0{\tily_{\ell-1,m}}

\def\Plp1{P_{\ell + 1}}
\def\Plm1{P_{\ell - 1}}

\def\llp1{\ell(\ell + 1)}
\def\tlp1{2\ell + 1}
\def\bh{{\bf h}}

\def\bk{{\bf k}}

\def\bs{{\bf s}}

\def\bsH{{\boldsymbol{H}}}

\def\bH{{\bf H}}

\def\ham{{\cal H}}

\def\bcJ{{\boldsymbol{\mathcal{J}}}}
\def\cF{{\mathcal{F}}}
\def\cJ{{\mathcal{J}}}
\def\cH{{\mathcal{H}}}
\def\cQ{{\mathcal{Q}}}
\def\cHs{\cH^{\mathrm{s}}}
\def\cHsp{\cH^{\mathrm{sp}}}
\def\itx{{\it x\/}\ }

\def\itz{{\it z\/}\ }

\def\itxz{{\it xz\/}\ }
\def\khat{{\bf{\hat k}}}

\def\xhat{{\bf{\hat x}}}
\def\yhat{{\bf{\hat y}}}
\def\zhat{{\bf{\hat z}}}

\def\tG{{\tilde G}}

\def\ket#1{|#1\rangle}
\def\bra#1{\langle#1|}

\def\mel#1#2#3{\langle#1|#2|#3\rangle}

\def\Fe8{Fe$_8$}
\def\Mn12{Mn$_{12}$}
\def\kh{{\hat k}}

\begin{document}

\title{Low-Temperature Phonoemissive Tunneling Rates in Single Molecule Magnets}
\author{Yun Liu}
\affiliation{University of Illinois, Department of Physics, 1110 W.\ Green St., Urbana, IL 61801}
\email[e-mail address: ]{yl2192@gmail.com}
\altaffiliation[Present address: ]{11-24 31st Ave., Apt. 9b, Long Island City, NY, 11106}
\author{Anupam Garg}
\affiliation{Northwestern University, Department of Physics and Astronomy, 2145 Sheridan Rd., Evanston, IL 60208}
\email[e-mail address: ]{agarg@northwestern.edu}
\date{\today}
\begin{abstract}
Tunneling between the two lowest energy levels of single molecule magnets with Ising type anisotropy, accompanied
by the emission or absorption of phonons, is considered. Quantitatively accurate calculations of the rates for such tunneling
are performed for a model Hamiltonian especially relevant to the best studied example, \Fe8. Two different methods are
used: high-order perturbation theory in the spin-phonon interaction and the non-Ising-symmetric parts of the spin Hamiltonian, and
a novel semiclassical approach based on spin-coherent-state-path-integral instantons. The methods are found to be in good
quantitative agreement with other, and consistent with previous approaches to the problem. The implications of these results
for magnetization of molecular solids of these molecules are discussed briefly.
\end{abstract}


\maketitle

\global\long\def\i{\mathrm{i}}
\global\long\def\e{\mathrm{e}}
\global\long\def\d{\mathrm{d}}

% Here starteth the paper.

\section{Introduction and Background}
\label{intro}

Single molecule magnets, also known as molecular nanomagnets, are a twenty-year old class of magnetic materials consisting of
organomagnetic molecules that form molecular solids. Their general characteristics are a total spin of about 10 per molecule
in the ground state, an absence of exchange interaction between different molecules, and magnetic anisotropy with an energy
difference per molecule of tens of Kelvin between easy and hard directions \cite{mol_mag_book,friedman_sarachik_2010}.
The most studied systems, \Fe8 and \Mn12, both have spin magnitude $j$ equal to 10, and uniaxial Ising-type anisotropy.
At low enough temperatures, 
only the $m = \pm 10$ Zeeman levels are occupied. On theoretical grounds, transitions between these two levels can only
take place via quantum tunneling, and for \Fe8 there is clear evidence that such tunneling really occurs
\cite{wernsdorfer-sessoli}, even though the tunneling matrix element is only $~10^{-8}\,$K or $1\,$peV in energy units.
However, quantum tunneling with significant probability transfer between the states involved can only happen if the states
are nearly degenerate in energy. If they are not, the tunneling degree of freedom (in this case the spin) must be coupled to
an environment which can supply or absorb the energy necessary to maintain energy conservation. This fact, coupled with the
general paucity of excitations with large energy, greatly restricts the dynamics of the {\it total\/} magnetization of the solid.
Several experiments~\cite{san97,ohm98,ohm-demag,wernsdorfer-ohm-prl,tho99,wer00,wer00b,tup02} find that magnetization relaxation
is slow, with non-exponential behaviour in time.

The current theoretical understanding of this slow relaxation \cite{pro96,avag09} is that interaction of the molecular spins with
the nuclear spins renders the quantum tunneling of the former incoherent, but because the nuclear spins that couple to a given
molecular spin can exchange only a rather limited amount of energy, the requirement of near-degeneracy of the Zeeman levels of the
molecular spin is weakened only moderately, and the two levels must lie within a narrow window of each other in order for transitions to
occur. Further relaxation can only take place due to the intermolecular dipole field, which can be quite inhomogeneous. If this field
happens to be such at a given molecular spin site as to bring that spin into near degeneracy, it will be able to flip. This flip will change
the field at other sites, potentially allowing those spins to relax. Monte Carlo and kinetic equation studies based on this model
have been done by several authors \cite{pro98ab,cuc99,fer0304,jv_book_chap9,vij13,lvg13},
all of whom obtain slow relaxation, and in some cases, an initial square-root time dependence as seen experimentally.

A central feature of the above model is that the transition rate between the $m = \pm j$ levels is insensitive to which one
is lower in energy. It thus allows the magnetization of a bulk sample to relax without relaxation of the energy, and the relaxation
is always toward the state of zero magnetization. As a result, this model cannot explain {\it magnetization\/} experiments in which
a magnetic field is applied to an initially demagnetized sample. In this case, it is essential to understand the relaxation of energy
as that is what drives the change in magnetization from zero to a nonzero value.

The obvious environment to which energy can be transferred is the phonons. The immediate puzzle is that the spin-phonon
interaction typically involves processes with $\Dta m = 1$ or $\Dta m =2$, while in the cases of \Fe8 and \Mn12, we require $\Dta m =20$.
Thus the relaxation must take place via a combination of spin tunneling and phonon emission. If we accept this hypothesis, the program
of understanding the magnetization experiments in molecular magnetic solids divides into two parts. The first part is to understand
the relaxation mechanism in a single molecule and calculate the relevant rate. The second part is to insert this rate into whatever
theory (for example, the kinetic equations) governs the dynamics of dipole-coupled molecules, and thus understand the behaviour of
the bulk solid. These two parts are logically separate and entail rather different ideas. The first part is entirely quantum
mechanical, while the second is entirely classical.

The purpose of the present paper is to address the first part. There are two prior calculations of phonoemisive
tunneling \cite{politi95,garg-adp}, but both are rather approximate, almost qualitative. Further, the first one is done for a
tetragonal spin anisotropy in the plane perpendicular to the primary (Ising) anisotropy axis, while the second one is done for
biaxial anisotropy. Further, it is not easy to estimate the error involved in these calculations. With this motivation, we
attempt in this paper to carry out a more accurate
treatment, making fewer approximations, and also trying to develop more than one method of calculation.

To explain more fully what we intend to do, let us consider the case of the \Fe8 molecule. Neglecting interaction with all other
degrees of freedom, the dynamics of the total spin of the molecule, $\bcJ$, are governed by the effective
Hamiltonian
\begin{equation}
\mathcal{H}^{\mathrm{s}}
   = -D\mathcal{J}_z^{2} + E(\mathcal{J}_x^{2}-\mathcal{J}_y^{2}) + C(\mathcal{J}_{+}^{4}+\mathcal{J}_{-}^{4})
        -g\mu_{B}\boldsymbol{H}\cdot\boldsymbol{\mathcal{J}}.
     \label{ham_fe8}
\end{equation}
Here, $\boldsymbol{\mathcal{J}}$ is the dimensionless spin angular momentum operator,
$\boldsymbol{H}$ is the external magnetic field, and $C$, $D$, and $E$ are anisotropy energies with $D > E \gg |C| > 0$.
The $g$-factor, $g$, is very close to 2, indicating an all-spin magnetic moment, and the magnitude of the total spin, $j$,
is 10 as already mentioned. The experimentally deduced values of $D$, $E$, and $C$ are 0.292, 0.046, and $2.9 \times 10^{-5}$ K,
respectively. Hence the molecule has an easy axis along $\zhat$ and hard axis along $\xhat$. The spin-phonon interaction can be
very generically written as
\beq
\mathcal{H}^{\mathrm{sp}}
 = \sum_{\mu} (F_{\mu}(\bcJ) a_{\mu} + F^{\dagger}_{\mu}(\bcJ) a^{\dagger}_{\mu}),
          \label{h_sp}
\eeq
where $a^{\dagger}_{\mu}$ and $a_{\mu}$ are phonon creation and annihilation operators for mode $\mu$, and $F_{\mu}(\bcJ)$ and
$F_{\mu}^{\dagger}(\bcJ)$ are functions of $\bcJ$ that have nonzero matrix elements with $\Dta m = 1$ and $\Dta m = 2$ (where
$m$ is the quantum number of $\mathcal{J}_z$). The phonons themselves are adequately treated in the harmonic approximation, i.e.,
with a Hamiltonian
\begin{equation}
  \mathcal{H}^{\mathrm{p}}=\sum_{\mu}\hbar\omega_{\mu}a_{\mu}^{\dagger}a_{\mu}.
\eeq

Let us first consider the single molecule magnet on its own, and restrict $\bsH$ to the \itx axis. Because the anisotropy ratios
$E/D$ and $C/D$ are small, it is reasonable to think of these terms, as well as the $H_x$ term, as perturbations, and work in the $\cJ_z$ basis.
The two lowest spin states, $m=-j$ and $m=j$, are then in resonance, and will
be split by quantum tunneling. One way to think of this tunneling is that it is caused by the three perturbing terms: $\mathcal{J}_x$
connects states with $\Delta m=\pm1$, $\mathcal{J}_x^{2}-\mathcal{J}_y^{2}$
connects states with $\Delta m=\pm2$, and $\mathcal{J}_{+}^{4}+\mathcal{J}_{-}^{4}$
connects states with $\Delta m=\pm4$. Various combinations of these three terms acting in sequence then connect the $m=-j$ and
$m=j$ states with each other, and split their degeneracy, i.e., give rise to tunneling. This method can be used for quantitative
calculations of tunnel splittings \cite{ks78,gara91,h-b96,jv_book_chap6}, and can even give accurate results \cite{park-garg-perturbative}
for the locations of the diabolical points where the splitting vanishes~\cite{wernsdorfer-sessoli,ag93}.

The same idea can be applied to find the matrix element for phonoemissive tunneling. Let us take $H_z > 0$ so that the $m=10$ state
is lower in energy than the $m=-10$ state. For the initial and final states, $\ket{i}$ and $\ket{f}$, let us take
\bea
\ket{i} &=& \ket{m = -10,\ {\mathrm{no\ phonons}}}, \nnu\\
\ket{f} &=& \ket{m = 10,\ {\mathrm{one\ phonon\ in\ mode\ }\mu(f)}}. 
\eea
We then consider as perturbations all possible strings of the $H_x$, $E$, and $C$ terms {\it plus\/} the spin-phonon interaction
$\mathcal{H}^{\mathrm{sp}}$ that take us from $\ket{i}$ to $\ket{f}$, with the restriction that the $m$ quantum number is always
increasing. The various strings, as well as the matrix elements of the associated spin operators, and the energy denominators,
are easily enumerated and calculated by symbolic manipulation programs. If the resulting matrix element is denoted by $\cF_{fi}$,
the rate or probability of transition per unit time is given by Fermi's golden rule as
\beq
\Gam = \frac{2\pi}{\hbar} \sum_f |\cF_{fi}|^2 \dta(E_f - E_i),
\eeq
where $E_i$ and $E_f$ give the energy of the magnetic molecule {\it plus\/} the phonons in the initial and final states.
It is of course the fact that the phonons have a continuum of energies that makes it possible to satisfy the delta function, and
thus obtain a nonzero transition rate.

We carry out this perturbative calculation in \Sno{sec:Perturbation-Theory}, with a generalization to nonzero temperature,
and including the case of absorption. The spin-phonon Hamiltonian and the phonon spectrum are discussed in more detail in this
section.

In \Sno{sec:Semiclassical-Calculation} we perform instead a semiclassical calculation based on spin coherent state path integrals.
This method has been very successfully applied to pure spin tunneling, but the application to an inelastic process involving
spin is completely novel as far as we know, and quite distinct from the classic calculations of dissipative tunneling involving
a continuous position-like degree of freedom~\cite{cal_leggett} in that for our problem tunneling would simply not take place without
the mechanism for inelasticity or dissipation. The path integral method is more compact than the perturbative calculation,
requires less numerical work, and yields a closed form answer in special cases. Finally, these two approaches provides checks
on each other.

The plan of the paper is as follows. In the next section we describe our theoretical model and some general features of the
phonoemissive and absorptive rates. Secs. \ref{sec:Perturbation-Theory} and \ref{sec:Semiclassical-Calculation} contain the
perturbative and semiclassical calculations. We apply the results of these calculations to \Fe8 in \Sno{sec:Results}, and
briefly discuss the implications of our findings for the magnetization problem described at the beginning of this section,
leaving a detailed examination of the issue for a later publication. Finally, technical aspects of the calculations are
explained in two appendices.

\section{General Aspects of the Problem}
\label{gen_aspects}

\subsection{The minimal theoretical model}
\label{ze_model}

If the magnetic molecule were truly isolated, and $H_z$ were zero, there would, as discussed in \Sno{intro}, be a 
tunnel splitting $\Delta$ between the two levels $m = \pm j$, and a state initially localized along one of the directions of
classical minima, say $m=-j$, would coherently flip-flop between $m=-j$ and $m=+j$, much like the inversion resonance in NH$_3$.
It is this tunnel splitting that is measured in Ref.~\cite{wernsdorfer-sessoli}, although the
transitions between the two levels are strongly incoherent in the real system, and there is no flip-flop. Instead, one-shot
transitions are induced by means of the Landau-Zener-Stuckelberg protocol, and the underlying matrix element is deduced from the
rate of relaxation of the net magnetization.

When $H_z$ is small but non-zero, there will be a relative bias $\varepsilon$ between the $+j$ and
$-j$ state, given by,
\begin{equation}
\varepsilon=E_{+j}-E_{-j}=-2jg\mu H_z.\label{eq:bias}
\end{equation}
The two level description between $-j$ and $+j$ is still accurate (provided that $H_z$ is not so large as to trigger the next
resonance, say between $m=-j$ and $m=j-1$), and an isolated molecule would still exhibit coherent oscillations with effective splitting
$(\Delta^{2}+\varepsilon^{2})^{1/2}$. In the real solid, interactions with other degrees of freedom destroy the coherence, and then,
as discussed in \Sno{intro}, transitions are in fact not possible for $\varepsilon \ne 0$ unless some environment soaks up or provides
the energy $\veps$. When there is a suitable environment, and the coupling to it is sufficiently strong, the transitions turn from
flip-flop to incoherent, with a probability that grows linearly with time. The coefficient of this growth is the rate that we seek.

In this paper, we shall take the spin Hamiltonian to be \eno{ham_fe8}, and restrict $\bsH$ to lie in the \itxz plane. Since the
interesting bias energies are of order 1 K, we need only consider long wavelength acoustic phonons. The mode label $\mu$ then
consists of the wavevector $\bk$ and a polarization index $s \in \{1, 2, 3\}$. We now need the spin-phonon interaction more explicitly
than \eno{h_sp}. The desired form can be obtained by promoting the classical magnetoelastic interaction~\cite{ll_v8}
to an operator, and takes the form
\begin{equation}
\mathcal{H}^{\mathrm{sp}}
  =\frac{1}{2}
     \sideset{}{'}\sum_{a,b,c,d=x}^{z}\Lambda_{abcd}(\partial_{a}u_{b}(\boldsymbol{0})+\partial_{b}u_{a}(\boldsymbol{0}))
          \{\mathcal{J}_{c},\mathcal{J}_{d}\},
     \label{def_Hsp}
\end{equation}
where $\Lambda_{abcd}$ is the magnetoelastic tensor, $\{\cdot,\cdot\}$ denotes the anticommutator, and $\boldsymbol{u}$ is the material
displacement field of the solid. The prime on the sum indicates the omission of certain terms from it, as we shall explain below.
Since we are limiting ourselves to acoustic phonons only, the field $\boldsymbol{u}(\boldsymbol{x})$ is given by~\cite{am_app_L}
\begin{equation}
\boldsymbol{u}(\boldsymbol{x})
  =\sum_{\mu} \sqrt{\frac{\hbar}{2m\omega_{\mu}N}} \boldsymbol{e}^{\mu}
        (a_{\mu}\e^{i\boldsymbol{k}^{\mu}\cdot\boldsymbol{x}}+a_{\mu}^{\dagger}\e^{-i\boldsymbol{k}^{\mu}\cdot\boldsymbol{x}}),
\end{equation}
where $m$ is the mass per unit cell of the lattice, $N$ is the number of unit cells in the solid, $a^{\dagger}_{\mu}$ and
$a_{\mu}$ are creation and annihilation operators for phonons of mode $\mu$, and $\boldsymbol{k}^{\mu}$ and
$\boldsymbol{e}^{\mu}$ are the wavevector and polarization of that mode.
The tensor components $\Lambda_{abcd}$ are not well known, so as a simplification we set~\cite{Lam_form}
\beq
\Lambda_{abcd} = \hf\Lambda (\delta_{ac}\delta_{bd} + \dta_{ad} \dta_{bc}).
\eeq
It is useful at this point to note that $\Lam$ has dimensions of energy, and that its
order of magnitude is expected to be comparable to $D$. With all the above simplifications, we get
\begin{equation}
\mathcal{H}^{\mathrm{sp}} =
    i \sum_{\mu} \lam_{\mu} (a_{\mu}-a_{\mu}^{\dagger})
             \sum_{a,b=x}^{z}(\kh_{a}^{\mu}e_{b}^{\mu}+\kh_{b}^{\mu}e_{a}^{\mu})
               \times \tshf \{\mathcal{J}_{a},\mathcal{J}_{b}\},
    \label{hsp_simple}
\end{equation}
where $\kh^{\mu} = k^{\mu}/|\bk^{\mu}|$, and we have introduced
\beq
\lam_{\mu} = 
    \Lam |\bk^{\mu}| \sqrt{\frac{\hbar}{2m\omega_{\mu}N}}
\eeq
to save writing. This quantity has dimensions of energy.

The phonon spectra for single molecule magnets are also not well known in general. But since only the low
frequency acoustic modes are of any relevance to us, in the same spirit that led us to simplify
$\Lam_{abcd}$, we take the phonons to be those of an isotropic continuous medium. That is, we take them to be exactly longitudinal
(L) and transverse (T) with the two transverse modes being degenerate for all directions of $\bk$, and to have a linear dispersion,
\beq
\om_L = c_L k, \quad \om_T = c_T k,
\eeq
with $k = |\bk|$, and $c_L$ and $c_T$ being the corresponding sound speeds. The sum over all phonon
modes can be replaced by an integral in the usual way,
\begin{equation}
\frac{1}{m N} \sum_{\mu} \to\frac{1}{8\pi^{3} \rho} \sum_{s = 1}^3 \int_{0}^{\infty}k^{2} dk\int_{\mathrm{S}^{2}} d^2\khat.
\end{equation}
Here, $\rho$ is the (mass) density of the material, $d^2\khat$ denotes an integral over all directions $\hat{\boldsymbol{k}}$,
and $s$ labels the polarizations (one longitudinal and two transverse).

As mentioned above, we omit certain terms from the sum in \eno{def_Hsp}. Here is why. The $m = \pm j$ states of the molecule  
are long-lived, and may be regarded as leading to quasi-equilibrium states of the solid as a whole. In particular, the solid should
have no strain in these states. This requirement, along with the linearity of $\cHsp$ in the strain, implies that $\mel{j}{\cHsp}{j}$
and $\mel{-j}{\cHsp}{-j}$ should vanish. The simplest way to ensure this is to omit the terms proportional to $\cJ_z^2$. The
alternative would be to project out just the $m = \pm j$ states, but the projection operator that achieves this is a high order
polynomial in  the components of $\bcJ$, and very difficult to work with. We believe that the differences between the two approaches
are physically insignificant, so we adopt the simpler one. Yet another alternative would be to keep the $\cJ_z^2$ terms but for
purposes of calculating $\Gam$, take the initial state of the lattice to be the strained one that results when the molecule is held
fixed in the $m=j$ or $m=-j$ state by an external agency. This approach also leads to a calculation that is harder but insignificantly
different in its physical implications from the ones that we perform.

\subsection{Interrelationships among various rates}
\label{symm_rates}

The type of question we wish to now calculate is the following. Suppose the spin is initially in the state $m=-j$, and the phonons
are at a temperature $T$. What is the probability that after a time $t$, the spin will be in the state $m=j$ irrespective of the
state of the phonons? This (inclusive) probability is given by
\begin{equation}
\mathcal{P}_{-j\to j}(t)
  = \sum_{\{n\}_{f}}\sum_{\{n\}_{i}}\frac{1}{\mathcal{Z}^{\mathrm{ph}}}\exp\bigl(-\beta E_{\{n\}_{i}}^{\mathrm{ph}}\bigr) \,
     |\mel{\{n\}_{f}, j}{\exp(-i t\mathcal{H}/\hbar)}{\{n\}_{i}, -j}|^{2},
    \label{P_inclusive}
\end{equation}
where $\{n\}_{i}$ and $\{n\}_{f}$ are the initial and final phonon configurations, $E_{\{n\}_{i}}^{\mathrm{ph}}$ is the energy of the
initial phonons, $\be = 1/k_B T$, and $\mathcal{Z}^{\mathrm{ph}}$ is the phonon partition function. Under the conditions discussed
in \Sno{intro}, except for very small $t$, this probability can be described in terms of a transition {\it rate\/}, $\Gam_{-j \to j}$.
That is,
\begin{equation}
\mathcal{P}_{- j\to j} = \Gam_{-j \to j}(\bH, T)\, t,
\eeq
and we have indicated that the rate depends on the field $\bH$ and the temperature. Under these same conditions, the rates for this
transition and its reverse are related by {\it detailed balance\/}, that is,
\beq
\Gam_{j \to -j}(\bH,T) = e^{\be\veps}\, \Gam_{-j \to j}(\bH,T), \label{det_bal}
\eeq
where the bias $\veps$ is defined by \eno{eq:bias}.
The content of \eno{det_bal} is better stated in words. It says that the ratio of the two rates is given by a Boltzmann factor
equal to the ratio of the probabilities of occurrence of the initial states in the Gibbs ensemble. In the mathematical form
(\ref{det_bal}) this content is all buried in plus and minus signs. When $H_z > 0$, the bias $\veps < 0$, the rate on the left side
involves phonon absorption, and that on the right involves phonon emission. When $H_z < 0$, the bias $\veps > 0$, and the left and right
sides of the equation pertain to phonon emission and absorption respectively. In all cases, the rate for phonon absorption is
$e^{-\be|\veps|}$ times that of emission.

While the property of detailed balance follows from very general considerations, we shall also see it explicitly in our
calculations in Secs.~\ref{sec:Perturbation-Theory} and \ref{sec:Semiclassical-Calculation}. We dwell on this point because in
the prior studies~\cite{pro96,avag09}, the
transition rates accompanied by energy transfer to or from the nuclear spins {\it violate\/} detailed
balance. Formally, this is because the nuclear spins are taken to be ``hot,'' or completely disordered. They are at effectively
infinite temperature, so any factor such as $e^{\be\veps}$ is equal to unity.

At this point, it pays to extract various factors on which these rates must depend on general grounds. First,
the relation (\ref{det_bal}) guarantees that their absolute temperature dependence is given by the Bose occupation factors,
\beq
\Gam_{+j \to -j}(\bH,T)
  \propto \frac{1}{e^{-\be\veps} - 1},  \quad
\Gam_{-j \to +j}(\bH,T)
  \propto -\frac{1}{e^{\be\veps} - 1},  \label{Gam_Bose}
\eeq
or, more explicitly, $(1 - e^{-\be|\veps|})^{-1}$ for emission, and $(e^{\be|\veps|} - 1)^{-1}$ for absorption. As $T \to 0$, these factors
tend to 1 and 0, respectively. Second, they must be proportional to the energy density of phonons (in the initial state for emission,
the final state for absorption) that can participate in the process, i.e., at the energy $|\veps|$.
This consideration gives an additional factor of
$|\veps|^3$. Third, to lowest order in the spin-phonon interaction, the rates must be proportional to $\Lam^2$. The remaining dependence
on the elastic properties of the solid can arise only via the parameters $\rho$ and $c$, the density and speed of sound, and dimensional
analysis then shows that it can only take the form
\beq
\Gam \sim \frac{\Lam^2}{\rho\hbar^4 c^5}
           \times \frac{\pm |\veps|^3}{e^{\mp\be\veps} - 1} \times G,
\eeq
where $G$ is a dimensionless quantity that we shall call the {\it spin tunneling cofactor\/}. It depends on $H_x$, $H_z$, and the
anisotropy ratios $E/D$ and $C/D$.
This general form may be seen in either Ref.~\cite{politi95} or Ref.~\cite{garg-adp}. More precisely,
we must replace the schematic factor $G/c^5$ by a weighted average over the longitudinal and transverse modes,
which the detailed calculations of the next two sections reveal to be of the form
\beq
\frac{2}{5}\frac{G_L}{c_L^5} + \frac{3}{5}\frac{G_T}{c_T^5}.
\eeq
We will choose the numerical factors so that final answer for $\Gam$ is written as
\beq
\Gam = \frac{\Lam^2}{2\pi\rho\hbar^4}
           \times \frac{\pm |\veps|^3}{e^{\mp\be\veps} - 1} 
            \times \left(\frac{G_L}{15 c_L^5} + \frac{G_T}{10 c_T^5}\right).
    \label{Gam_std_form}
\eeq

To differentiate the various rates, let us add to $\Gam$ and $G$, superscripts ``abs'' and ``em'' to indicate absorption and
emission, and subscripts $-j \to j$ or $j \to -j$ to show the direction of transition. There are four
$\Gam$'s to be considered, which obey various relationships, better expressed in terms of the $G$'s. Two of these
are just reexpressions of detailed balance:
\begin{equation}
\begin{aligned}
G_{-j\to+j}^{\mathrm{abs}}(\varepsilon)
  & = G_{+j\to-j}^{\mathrm{em}}(\varepsilon), &  &  &
G_{+j\to-j}^{\mathrm{abs}}(\varepsilon)
  & = G_{-j\to+j}^{\mathrm{em}}(\varepsilon).
\end{aligned}
\label{eq:observation-1}
\end{equation}
Two other relationships follow from time-reversal symmetry combined with a 180$^{\circ}$ rotation about the \itz axis,
i.e., the invariance of the system under $m\to-m$ and $H_z\to -H_z$. These are,
\begin{equation}
\begin{aligned}
G_{-j\to+j}^{\mathrm{abs}}(\varepsilon)
  & =G_{+j\to-j}^{\mathrm{abs}}(-\varepsilon), &  &  &
G_{-j\to+j}^{\mathrm{em}}(\varepsilon)
  & =G_{+j\to-j}^{\mathrm{em}}(-\varepsilon).
\end{aligned}
\label{eq:observation-2}
\end{equation}
It must be remembered that these relations hold separately for $G_L$ and $G_T$. For the rates, \eno{eq:observation-2} implies that,
\beq
\Gamma^{\rm abs/em}_{-j\to j}(\varepsilon)
  = \Gamma^{\rm abs/em}_{j\to-j}(-\varepsilon),
     \label{t_reverse}
\eeq

A very similar argument shows that the $\Dta m = 1$ and $\Dta m =2$ amplitudes will be odd and even functions respectively of $H_x$
if $j$ is an integer, and the other way around if $j$ is a half-integer. This property is a useful calculational check.

There is one more property of the rates that may be described as a symmetry. Every rate ends up being a sum of two subrates,
one involving spin-phonon processes with $\Dta m =1$, and the other involving $\Dta m =2$. That this is so, and that there is no
interference between these two alternatives is because our spin-phonon interaction is invariant under {\it all\/} rotations,
not just the restricted rotations in the point group that would be relevant for a real solid. Angular momentum is thus fully
conserved, and so in principle we
can distinguish tunneling that required the $\Dta m =1$ vs.\ the $\Dta m=2$ part of $\cHsp$ by looking at the angular momentum carried
by the emitted or absorbed phonon. Since the remaining change in $m$ (19 or 18) must then be due to the action of the spin Hamiltonian
$\cHs$ alone, the corresponding amplitudes must be odd and even in $H_z$, respectively. (The subrates
themselves are of course always even, but this fact is another useful check on the calculations.) This point is perhaps clearer in the
context of the calculations themselves, but it is worth noting here also.
Naturally, this property will not hold strictly for a real system, but it may still be approximately true.

Because of all these symmetries, we need only give four cofactors ($G_L$ and $G_T$, each for $\Dta m =1$
and $\Dta m=2$) for anyone of the four rates. The corresponding quantities for the other rates then follow from the symmetry relations,
and there is no need to give them separately. For our standard process, we shall take $-j \to j$ phonoemission, and give the $G's$
for it. Further, to avoid cluttering the formulas, we shall omit the distinguishing suffixes wherever we can do so without creating
ambiguity. Thermostatistical factors such as $\veps^3/(1 - e^{-\be\veps})$, as well as the dimensional factors involving $\Lam^2$
and the sound speeds can be incorporated for any other process as needed.

\section{Perturbation Theory}
\label{sec:Perturbation-Theory}

The evaluation of $\mathcal{P}_{-j\to j}$ [see \eno{P_inclusive}] requires a generalization of the second-order Fermi golden rule~\cite{baym},
which is achieved as follows. We divide the total Hamiltonian for the molecule and the phonons into two parts,
$\mathcal{H}_0$, and $V$, where
\beq
\mathcal{H}_0 = -D\cJ_z^2 - g\mu_B H_z \cJ_z + \mathcal{H}^{\mathrm{p}},
\eeq
and
\beq
V = \cH_1 + \cH_2 + \cH_4 + \cHsp,
\eeq
with  
\begin{equation}
\mathcal{H}_{1}  =-g\mu H_x\mathcal{J}_x, \qquad \mathcal{H}_{2} = E(\mathcal{J}_x^{2}-\mathcal{J}_y^{2}), \qquad
   \mathcal{H}_{4} = C(\mathcal{J}_{+}^{4} + \mathcal{J}_{-}^{4}).
\end{equation}
The usual perturbative expansion gives us
\beq
e^{-i \mathcal{H}t/\hbar}
 =  e^{-i \mathcal{H}_0t/\hbar}
  \Bigl[ 1 + \frac{1}{i\hbar} \int_0^t dt_1 V(t_1)
           + \frac{1}{(i\hbar)^2} \int_0^t dt_1 \int_0^{t_1} dt_2 V(t_1) V(t_2) + \cdots \Bigr],
     \label{pert_expans}
\eeq
with $V(t)$ being the perturbation in the interaction picture. We must take the matrix element of this operator between the initial
and final states, which we simplify as follows. First, our final state differs from the initial one by $\Dta m = 2j$, which
is 20 for \Fe8. To get such a large change in $m$, one must go to an order such that there are sufficiently many interaction terms
$V(t_i)$. There are many ways to achieve this. For example, in sixth order, we could select the following sequences of terms:
\beq
\cH_4 (t_6), \cH_4 (t_5), \cH_2(t_4), \cH_4(t_3), \cHsp(t_2), \cH_4(t_1);   \label{seq1}
\eeq
and in seventh order we could select
\beq
\cH_2(t_7), \cHsp (t_6), \cH_4 (t_5), \cH_4(t_4), \cH_4(t_3), \cH_4(t_2), \cH_1(t_1). \label{seq2}
\eeq
It is evident that each sequence can be considered to correspond to a discrete path in the space of Zeeman states.
All possible paths must be considered, and their contribution to the transition matrix element must be added together. To make
the calculation tractable, we make the following simplifications. First, we can divide the paths into different types of classes
characterized by how many times each one of the four parts of $V(t)$ appears. We demand that all classes must be considered subject
to the constraint that the $m$ quantum number must be strictly increasing (or strictly decreasing for the $j \to -j$ transition).
Thus the sequence
\beq
\cH_2(t_8), \cH_2(t_7), \cHsp (t_6), \cH_4 (t_5), \cH_4(t_4), \cH_4(t_3), \cH_4(t_2), \cH_1(t_1), \label{seq3}
\eeq
will be ignored since when we compare it with (\ref{seq2}), it is seen to require a step in which $m$ decreases~\cite{dps}.
Second, we demand that in each path, $\cHsp$ appear once and only once. If it does not appear at all, we can not allow for energy
conservation if there is a significant bias, and so the path in question cannot contribute to the incoherent transition rate. (It is
instead, part of the contribution to $\Dta$, the coherent flip-flop tunnel splitting.) And, we limit it to only one appearance
because $\Lam$ is small, so it suffices to work to lowest order in $\Lam$. This is another way of saying that we limit ourselves to
one-phonon processes. In the transition matrix element, therefore, we need only display the phonon occupation number of the mode that is
affected, and the transition probability simplifies to
\begin{equation}
\begin{aligned}
\mathcal{P}_{-j \to j}^{\mathrm{one-phonon}}
    & =\sum_{\mu}\sum_{n_{\mu}}\frac{\mathrm{e}^{-\beta\hbar\omega_{\mu}n_{\mu}}}{\mathcal{Z}_{\mu}^{\mathrm{ph}}}
           \lvert\bra{n_{\mu}+1, j}\exp\{-i t\mathcal{H}/\hbar\}\ket{n_{\mu}, -j}\rvert^{2}\\
    & \quad+\sum_{\mu}\sum_{n_{\mu}}\frac{\mathrm{e}^{-\beta\hbar\omega_{\mu}n_{\mu}}}{\mathcal{Z}_{\mu}^{\mathrm{ph}}}
            \lvert\bra{n_{\mu}-1, j}\exp\{-i t\mathcal{H}/\hbar\}\ket{n_{\mu}, -j}\rvert^{2},
\end{aligned}
\label{eq:tunneling-approximation}
\end{equation}
where we have explicitly separated the phonoemissive and phonoabsorptive processes, and it is understood that we employ the
perturbation expansion (\ref{pert_expans}) with the simplifications already mentioned.

We now note that since $\cHsp$ is to appear only once in the expansion of $e^{-i\cH t/\hbar}$, the phonon part of the transition
matrix element is always
\begin{equation}
   \bra{n_{\mu}+1}a_{\mu}^{\dagger}\ket{n_{\mu}} =\sqrt{n_{\mu}+1},
   \qquad {\mathrm{or}}\qquad
     \bra{n_{\mu}-1}a_{\mu}\ket{n_{\mu}} =\sqrt{n_{\mu}},
\end{equation}
for emission and absorption respectively. Summing over the Boltzmann weight then gives precisely the Bose thermal
occupation numbers $\left\langle n_{\mu}+1\right\rangle =(1-\exp\{-\beta\hbar\omega_{\mu}\})^{-1}$
and $\left\langle n_{\mu}\right\rangle =(\exp\{\beta\hbar\omega_{\mu}\}-1)^{-1}$ for these two processes, and their rates are guaranteed
to obey detailed balance. Henceforth we consider only $-j \to j$ phonoemission; this implies that $\veps < 0$. We also omit the
Bose factors which can be restored at the end. Formally this is like working at $T=0$.

The remaining $V(t_i)$ and $\cHsp$ in any path give rise to a number of factors of the form $e^{-i(E_k - E_{\ell}) t_i}$, where $E_k$
and $E_{\ell}$ are the energies of intermediate states along the path. When we integrate over all the $t_i$, all but one of these
integrations will generate energy denominators of the form $(E_k - E_i)$, and the remaining one will generate an overall sinc function
which upon squaring can be replaced by $t \dta(E_f -E_i)$ by standard arguments~\cite{baym}. The upshot is that 
\begin{equation}
\mathcal{P}^{\rm em}_{- j\to j} = \Gam^{\rm em}_{-j \to j}\, t,
\eeq
where
\begin{equation}
\Gamma^{\rm em}_{- j\to j}
   =  \frac{2\pi}{\hbar} 
      \sum_{\mu}\delta(\varepsilon+\hbar\omega_{\mu})
        \lvert\mathcal{F}_{- j\to j}^{\mu,\mathrm{em}}\rvert^{2}.
                  \label{eq:transition-rate-1}
\end{equation}
The quantity $\cF$ (suppressing the suffixes) is a transition matrix element with the following structure:
\begin{equation}
\mathcal{F}
  = \sum \frac{\bra{f}V\ket{s_{n-1}}\bra{s_{n-1}}V\ket{s_{n}}\cdots\bra{s_1}V\ket{i}}
              { (E_i - E_{n-1})(E_i - E_{n-2})\cdots(E_i - E_1)}.
\end{equation}
Here the $s_k$'s denote intermediate states, with energies $E_k$ (which include the energies of the phonons), the $V$'s are the
interactions in the usual Schrodinger picture (which are time independent), and the sum is over all paths from $-j$ to $j$ with the
restrictions mentioned above.
For the numerator, we note that since $m$ is always increasing along a path, the product of spin matrix elements will
always contain the factor
\beq
\mel{j}{\cJ_+^{2j}}{-j} = (2j)!.   \label{num_factor}
\eeq
In the denominator, the energies $E_k$ include those of the phonons.

The problem of calculating $\cF$ thus reduces to enumerating all the (restricted) paths, multiplying the correct
number of relevant couplings, $E$, $C$, and $H_x$ in the numerator (there is always one factor of $\Lam$), the energy differences
in the denominator, and then summing over paths. This task is easily automated to a computer, but before doing that it is worth
simplifying it further by anticipating the structure of the sum over phonon modes. Let us define
\beq
\cQ_{ab} = \hf \{\mathcal{J}_{a},\mathcal{J}_{b}\}.   \label{def_qab}
\eeq
There are six such operators in all, but as explained in \Sno{ze_model}, $\cHsp$ does not contain $\cQ_{zz}$. Even if such
a term were present in $\cHsp$, however, it would lead to $\Delta m=0$ processes, so we would discard it in accord with the
requirement that $m$ be strictly increasing along a path. For the same reason, we can make the following replacements for
the $\Dta m = 2$ operators:
\beq
\cQ_{xx} \to \frac{\cJ_+^2}{4}, \quad
\cQ_{xy} \to -i\frac{\cJ_+^2}{4}, \quad
\cQ_{yy} \to -\frac{\cJ_+^2}{4}.
\eeq
(We show the replacements for the $-j$ to $j$ rate; for $j$ to $-j$, we simply use the Hermitean conjugates.)
Further, since the $\cJ_+^2$ operator is incorporated in the common factor (\ref{num_factor}), we need only keep track of the numerical
factors of $1/4$, $-i/4$, and $-1/4$. Matters are slightly more involved for the $\Dta m = 1$ operators. This time, we have to keep track
of where in the $-j$ to $j$ chain they act. It is easy to see that the replacement rule is
\beq
\mel{m+1}{\cQ_{xz}}{m} \to \tshf (m + \tshf), \quad
\mel{m+1}{\cQ_{yz}}{m} \to -\tfrac{i}{2} (m + \tshf),
\eeq
the $\cJ_+$ part being already absorbed in the factor (\ref{num_factor}). In fact, if we change the common factor to
\beq
\mel{j}{\bigl(\tshf\cJ_+\bigr)^{2j}}{-j} = \frac{(2j)!}{2^{2j}},   \label{num_factor2}
\eeq
we can use the simpler replacement rules
\beq
(\cQ_{xx}, \cQ_{xy}, \cQ_{yy}, \cQ_{xz}, \cQ_{yz})
  \to \bigl(1, -i, -1, (m + \tshf), -i(m + \tshf) \bigr).
\eeq

It follows that we can split $\mathcal{F}$ into two sums, corresponding to the value of $\Dta m$ carried by $\cHsp$. 
In other words, we may write,
\beq
\cF^{\bk s}
   = -i \lam_{\bk s} \bigl[\kh_- e^{\bk s}_- M_2(\om_{\bk s}) + (\kh_- e^{\bk s}_z + \kh_z e^{\bk s}_-) M_1(\om_{\bk s}) \bigr],
   \label{F_structure}
\eeq
where,
\beq
\kh_{\pm} = \kh_x \pm i \kh_y, \quad e^{\bk s}_{\pm} = e^{\bk s}_x \pm i e^{\bk s}_y,
\eeq
and $M_{m}(\om_{\bk s})$ is the rest of the matrix element, with the subscript showing the value of $\Dta m$. This matrix element is
dimensionless because we have factored out $\lam_{\bk s}$. We have
written out the phonon mode index $\mu$ more fully to show that it consists of the wavevector $\bk$ and polarization label $s$.
For $m=2$, the matrix elements $M_m$ that remain at this stage have the structure
\beq
M_2 = 2\frac{(2j)!}{2^{2j}} \sum_{\rm paths} \frac{\rm product\ of\ coupling\ constants}{\rm product\ of\ energy\ denominators},
\eeq
where the additional factor of 2 arises from the symmetry of the tensor $(\kh^{\mu}_a e^{\mu}_b + \kh^{\mu}_b e^{\mu}_a)$,
the ``coupling constants'' are $-g\mu_B H_x$, $2E$, and $16 C$. Note that the coupling constant $\lam_{\bk s}$ from $\cHsp$
has already been removed from the numerator. For $M_1$, there is an additional factor of $(m+\hf)$ in the numerator, where $m$ depends
on where in the path $\cHsp$ acts.

As part of the sum over all phonon modes, we must sum over the three polarizations, and integrate over the directions of $\bk$. We
show how this is done in Appendix \ref{pol_avg}. The result has two contributions, one from the longitudinal modes, and the other from the two
transverse modes taken together. We may write this as
\beq
\int d^2{\hat\bk} \sum_s |\cF^{\bk s}|^2  = F_L(k) + F_T (k),
\eeq
where
\beq
F_{\al}(k) 
  = \frac{16\pi}{15} |\lam_{k\al}|^2 \, \Bigl[ |M_1(c_{\al} k)|^2 +  |M_2 (c_{\al} k)|^2 \Bigr]
      \times \begin{cases}
                  2, & \al = L, \cr
                  3, & \al = T. \cr
              \end{cases}
   \label{FL_FT}
\eeq
It remains to integrate over the magnitude of the phonon wavevector. This is trivially done because of the energy
conserving delta function. Furthermore, $c_{\al} k$ is replaced by $|\varepsilon|/\hbar$ for both the $L$ and $T$ modes,
so if we write the $M$'s in terms of the spin anisotropies, $D$, $C$, and $E$, the external field $H_x$, and the bias
$\varepsilon$ (as a proxy for $H_z$), we get the same function for both the $L$ and $T$ modes.
Hence, we obtain
\begin{equation}
\Gamma^{\rm em}_{- j\to j}
   =\frac{4\Lambda^{2}}{\pi\rho\hbar^{4}}
        \left(\frac{1}{15c_{L}^{5}}+\frac{1}{10c_{T}^{5}}\right) |\veps|^3
                \left(\lvert M_{1, - j\to j}^{\mathrm{em}}\rvert^{2}
                        +\lvert M_{2, - j\to j}^{\mathrm{em}}\rvert^{2}\right),
    \label{Gam_pert}
\end{equation}
where the suffixes on the $M$'s now indicate the value of $\Dta m$ as well as the transition involved more fully.
In terms of the spin tunneling copfactors $G_L$ and $G_T$ introduced in \Sno{symm_rates}, \eno{Gam_pert} means that
\begin{equation}
G_L^{\rm pt} = G_T^{\rm pt}
  = 8 \bigl[\lvert M_{1, -j\to j}^{\mathrm{em}}\rvert^{2}
      +      \lvert M_{2, -j\to j}^{\mathrm{em}}\rvert^{2} \bigr],
\end{equation}
where the superscript `pt' stands for `perturbation theory,' and we may refer to both $G_L^{\rm pt}$ and $G_T^{\rm pt}$ by just
one name, $G^{\rm pt}$, because of their equality. The power of five to which the speeds of sound occur is a major source
of uncertainty for the transition rate; and since $c_{T}$ is typically
half of $c_{L}$, the transverse contribution is likely to dominate.

Before moving onto the semiclassical calculation, let us see how the general consequences of time-reversal symmetry that were
mentioned in \Sno{symm_rates} play out in the above calculation. Under
the transformation $m\to-m$ and $H_z\to -H_z$, the energy denominators remain
unchanged, while for the numerators, those for paths containing a phonon-coupling
term with $\Delta m=\pm1$ will change sign, but those with $\Delta m=\pm2$ will not. As a result,
\begin{equation}
\begin{aligned}
M_{1, -j\to j}^{\mathrm{em}}(\varepsilon) & = - M_{1, -j\to j}^{\mathrm{em}}(-\varepsilon), &  &  &
M_{2, -j\to j}^{\mathrm{em}}(\varepsilon) & = M_{2, -j\to j}^{\mathrm{em}}(-\varepsilon).
\end{aligned}
\end{equation}
After taking the squared absolute values, the minus sign in the $M_1$ relation is immaterial, so 
\eno{t_reverse} follows. Furthermore we see that $M_1(0) = 0$ due to this antisymmetry.

In the same vein, we look at $M_1$ and $M_2$ as a function of $H_x$. When $j$ is an integer the transition from
$-j$ to $+j$ (and vice versa) must take an even number of steps in $m$. Since both $\mathcal{H}_{2}$ and $\mathcal{H}_{4}$ have
even $\Delta m$, the remaining combination of $\mathcal{H}_{1}$ (through which the external field $H_x$ appears) and $\cQ_{ab}$
must also result in an even $\Delta m$. Therefore, the $M_1$ term, which results from the $\Dta m =1$ part of $\cQ_{ab}$,
must include an odd number of appearances of $\cQ_1$, and so must be
an odd function of $H_x$. In the same way, the $M_2$
term is an even function of $H_x$. Consequently, at $H_x=0$ the $M_1$
contribution also vanishes. We shall find the same thing in the semiclassical calculation.

\section{Semiclassical Instanton Calculation}
\label{sec:Semiclassical-Calculation}

Our goal in this section is to treat the spin by semiclassical methods, exploiting the fact that in many
interesting cases, the spin is large. This is formally a different approximation from perturbation theory, although
when $j$ is large, the large number of intermediate states leads to a certain similarity. Another reason for the semiclassical
approach is that it leads to a compact answer for the rate where the perturbative one does not.

The semiclassical approach we use will be based on spin coherent state path integrals. In order not to interrupt their application
to the problem of this paper, we present a brief review of their use in the context of pure spin tunneling, especially for \Fe8.
We then apply them to the phonoemission and absorption, and then discuss these phenomena in the specific case of \Fe8.

\subsection{Pure spin tunneling via instantons}
\label{instantons_fe8}

Let us consider the pure spin (no phonons) tunneling amplitude
\beq
U_{fi}(t) = \mel{j}{\exp(-i\cHs_0 t/\hbar)}{-j},
\end{equation}
where $\cHs_0$ is the spin Hamiltonian (\ref{ham_fe8}) with $H_z=0$:
\begin{equation}
\mathcal{H}_0^{\mathrm{s}}
   = -D\mathcal{J}_z^{2} + E(\mathcal{J}_x^{2}-\mathcal{J}_y^{2}) + C(\mathcal{J}_{+}^{4}+\mathcal{J}_{-}^{4})
        -g\mu_{B} H_x \mathcal{J}_x.
      \label{ham_dps}
\end{equation}
Since we intend to evaluate it via spin coherent state path integrals, we give our definition of spin coherent states, which is
\begin{equation}
\ket{\bar{z}} = (1+z\bar{z})^{-j}\exp\{\bar{z}J_{+}\}\ket{j, -j}, \qquad
\bra{z} = (1+z\bar{z})^{-j} \bra{j, -j}\exp\{zJ_{-}\}.
\end{equation}
Here, $z$ is a complex number giving the coordinates of the state in the spin phase space $\mathrm{S}^{2}$
expressed via stereographic projection as $\mathbb{C}+``\infty"$. As before, an overbar denotes complex conjugation.
The preimage of this projection on the unit sphere is a vector $\bs$ with Cartesian components
\begin{equation}
\begin{aligned}
s_x & =\frac{z+\bar{z}}{z\bar{z}+1}, &  &  &
s_y & =\frac{1}{i}\frac{z-\bar{z}}{z\bar{z}+1}, &  &  &
s_z & =\frac{z\bar{z}-1}{z\bar{z}+1}.
\end{aligned}
\end{equation}
The vector $\bs$ gives the direction in which the spin ``points'' when it is in the state $\ket{\bar z}$.
In particular, the spin state $m=-j$ corresponds to $\bar{z}=0$ and $\bs = -\zhat$, while $m=+j$ corresponds to the point at infinity
and $\bs= \zhat$. In the classical limit, the operator $\bcJ$ has vanishing relative fluctuation about its mean value, $j\bs$.
We can see this from the matrix elements,
\begin{equation}
\begin{aligned}
\bra{z}\mathcal{J}_{a}\ket{{\bar z}} & =js_{a}(z), &  &  &
\tfrac{1}{2}\bra{z}\left\{ \mathcal{J}_{a},\mathcal{J}_{b}\right\} \ket{{\bar z}}
   & =j(j-\tfrac{1}{2})s_{a}(z)s_{b}(z)+\tfrac{1}{2} j\delta_{ab}.
\end{aligned}
\end{equation}

In what follows, it will help to think of $0$ and $t$ as ``initial'' and ``final'' times. We therefore define
\beq
t_i \equiv 0, \quad t_f \equiv t.
\eeq
With this relabelling, the path integral expression for $U_{fi}$ is
\beq
U_{fi}(t_f,t_i)
   =\mathcal{N}\int\mathcal{D}z\, \exp\left(iS^{\mathrm{spin}}[z]/\hbar \right). \label{ufi}
\eeq
Here, the sum is over all paths that go from ${\bar z}_i  = {\bar z}(t_i) = 0$ to $z_f = z(t_f) = \infty$,
$S^{\mathrm{spin}}$ is the semiclassical action for the spin Hamiltonian $\cHs_0$, and $\mathcal{N}$ is a normalization constant.
(The exact forms of $\mathcal{N}$ and $S^{\mathrm{spin}}$ will not be needed; for complete details see \cite{garg-stone-instanton}.)
Since $\cHs_0$ is a matrix of finite order, $2j+1$, the amplitude $U_{fi}$ is an entire function of $t_f$ and $t_i$, viewing these
as complex variables. The instanton method is essentially an infinite-dimensional steepest descent approximation to $U_{fi}$'s
analytic continuation onto the imaginary time axis,
\beq
t_f \to -i\tau_f, \quad t_i \to -i\tau_i.
\eeq
The instanton is a path, $\{z_d(\tau), {\bar z}_d(\tau)\}$, or equivalently, $\bs_d(\tau)$ (the subscript d stands for dominant),
that obeys the boundary conditions and stationarizes the action $S^{\mathrm{spin}}$. As in the ordinary steepest descents
method, we restrict the paths over which we integrate to small fluctuations around the instanton,
\beq
\xi(\tau) = z(\tau) - z_d(\tau), \quad
{\bar\xi}(\tau) = {\bar z}(\tau) - {\bar z}_d(\tau),
\eeq
and, at the same time, expand the action to quadratic order in the fluctuations,
\beq
S^{\rm{spin}} = S[z_d, {\bar z}_d] + \dta^2 S(\xi,{\bar\xi}).
\eeq
(We write the quadratic term as $\dta^2 S$ because it can be thought of as the second order change in $S$ when the path is varied from
$z_d, {\bar z}_d$ to $z_d + \xi, {\bar z}_d + {\bar\xi}$; the first order change vanishes on account of stationarity.)
Thus, the sum over all paths is replaced by one over small fluctuations about the instanton:
\begin{equation}
\mathcal{N} \int\mathcal{D}z \exp(-S^{\rm spin}/\hbar)
   \to e^{-S[z_d, {\bar z}_d]/\hbar} \times \mathcal{N} \int\mathcal{D}\xi e^{-\dta^2 S(\xi, {\bar\xi})/\hbar}.
      \label{small_flucts}
\end{equation}
(The action term in the exponents has been changed from $iS$ to $-S$ because of the continuation to imaginary or
Euclidean times. This is further explained below.)

In many problems it happens that there is more than one stationarizing path or instanton. In such cases, the right hand
side of \eno{small_flucts} must be a sum of similar terms, one for each instanton. This is precisely the situation that
arises with the Hamiltonian (\ref{ham_dps}). When $C=0$, there are two instantons that wind about the hard axis in opposite
directions, and when $C \ne 0$, no matter how small, there are two additional instantons.
For example, for $C = 0$ and $H_x = 0$ the instanton paths are~\cite{garg-stone-instanton}
\begin{equation}
\bs_d(\tau;q,\tau_{c})
  = i q\sqrt{\frac{D-E}{2E}} \mathrm{sech}\,\Omega(\tau - \tau_{c})\, \xhat 
       + q\sqrt{\frac{D+E}{2E}} \sech\,\Omega(\tau - \tau_{c})\, \yhat
       + \tanh\Omega(\tau - \tau_{c})\, \zhat.
    \label{inst_fe8}
\end{equation}
Here, $\Omega=2j\sqrt{D^{2}-E^{2}}/\hbar$ is the characteristic instanton frequency, and $q = \pm1$ is a parameter
that distingusihes the two windings. The (imaginary time, or Euclidean) action for these two windings has the same real
part, but different imaginary parts,
\beq
\frac{1}{\hbar} S[z_d(\tau), {\bar z}_d(\tau)] = I + i q\frac{\Tta}{2},
\eeq
and it is interference between them that is responsible for the experimentally observed diabolical
points~\cite{wernsdorfer-sessoli,ag93}. (For the case $E=0$ and $C\ne 0$, all four instantons
interfere~\cite{park-garg-perturbative}.)

The parameter $\tau_c$ in \eno{inst_fe8} is the center of the instanton where it switches between its end point values,
$\bs_d(\tau_i) = -\zhat$ and $\bs_d(\tau_f) = +\zhat$. To understand its importance, we note that the frequency $\Om$
is the inverse of time scale over which the instanton makes this switch. This time scale is very short because
quite generally, $\hbar\Om$ is comparable to the energy difference between the $m=j$ and $j-1$ energy levels, which is
very large compared to $\Dta$ and $\eps$, the scales of interest to us. The corresponding condition on the time scales
with which are concerned is 
\beq
\tau_f - \tau_i \gg \Om^{-1}.   \label{inst_ok}
\eeq
Strictly speaking, the instanton is only defined for $\tau_f \to +\infty$, $\tau_i \to -\infty$, but in fact, we may
employ steepest descents as long as condition (\ref{inst_ok}) holds. From the time-translation invariance of the
Hamiltonian, it is then apparaent that $\tau_{c}$, the center of the instanton, can be located essentially anywhere in
the interval $(\tau_i, \tau_f)$ without affecting the value of the action. This is a general feature of all instanton
solutions, and it means that when we sum over all fluctuations about the instanton, fluctuations that are equivalent to
a shift in the center time, $\tau_c$, have a special role, and must be summed over separately. In other words,
\begin{equation}
\mathcal{N} \int\mathcal{D}\xi
   \to \int_{\tau_i}^{\tau_{f}}\!\d\tau_{c} \times \mathcal{N}' \int'\mathcal{D}\xi,
\end{equation}
where $\mathcal{N}'$ is another normalization constant (which we shall also not need), and the prime on the integral
means that fluctuations equivalent to a time translation are excluded.

Let us temporarily specialize to the case $C = 0$. Then there are only two types of instantons, with $q = \pm 1$,
and the transition amplitude is given by
\beq
U_{fi}
 = \sum_{q=\pm 1} \int_{\tau_i}^{\tau_{f}}\!\d \tau_{c}\, e^{- (I + iq \Tta/2)/\hbar}
      \times \mathcal{N}' \int'\mathcal{D}\xi\, e^{-\dta^2 S/\hbar}.   \label{ufi_steep}
\eeq
The integral over $\tau_c$ yields a factor of $\tau_f - \tau_i$.
The Gaussian fluctuation integral is equal for both values of $q$ because of symmetry. Hence, writing
\beq
e^{-I} \times \mathcal{N}' \int'\mathcal{D}\xi\, e^{-\dta^2 S/\hbar} = \frac{W}{4\hbar}, \label{gauss_fluct}
\eeq
and
\beq
\Dta = W \cos(\tshf\Tta), \label{def_W}
\eeq
we obtain
\beq
U_{fi} = \frac{\Dta}{2\hbar}(\tau_f - \tau_i). \label{U_one_inst}
\eeq

Equation (\ref{U_one_inst}) appears to violate unitarity. However, because $\Om^{-1} \ll \tau_f - \tau_i$, one must
consider multi-instanton paths. As long as the centers of these instantons are well separated on the $\Om^{-1}$ time scale,
these paths also stationarize the action, and are valid saddle points. When these paths are included, terms of higher
order in $(\tau_f - \tau_i)$ are generated~\cite{coleman77}, and \eno{U_one_inst} is seen to be the first term in the
expansion of $\sinh (\Dta(\tau_f - \tau_i)/2\hbar)$. Since the structure of this expansion is fairly straightforward, it
generally suffices to consider only one-instanton paths to see it. The analytic continuation back to the real time axis
yields $\sin(\Dta t/2\hbar)$ for $U_{fi}$, the expected answer for a system with two ground states tunnel split by an energy $\Dta$.

If we now let $H_x$ become nonzero while keeping $C = 0$, the above picture holds unchanged, except that $I$ and $\Tta$ are functions
of $H_x$. The splitting vanishes whenever $\Tta$ is an odd multiple of $\pi$, producing a diabolical point. 

Next let us ask what happens if we allow $C$ to be nonzero. In this case, everything said above continues to be true for $H_x = 0$
and sufficiently small $H_x$. The reason is that the two new instanton solutions which also stationarize $S$ have very large
action, so they are physically inconsequential. This ceases to be true once $H_x$ crosses a certain value (depending on how large
$C$ is). Then, one of the new instantons has the least action, and since it is without an interfering partner, the diabolical points
move off the $H_x$ axis into the $H_x$-$H_y$ plane~\cite{garg-kececioglu-jump,bruno06,li-garg}.

We shall confine our analysis of phonoemissive tunneling to small value of $H_x$, where the extra $C\ne 0$ instantons are not
important. As far as we are aware, there are no systematic studies of the magnetization process as a function of $H_x$.

\subsection{Application to phonoemissive tunneling}
\label{phono_instanton}

For the reasons given in \Sno{symm_rates}, we need calculate only the $-j \to j$ phonoemissive rate at $T=0$. Answers for $T \ne 0$
merely require inclusion of Bose occupation factors. We are thus led to consider the one-phonon transition matrix element
\beq
A^{\mu}_{fi}(t) = \mel{j, n_{\mu} =1}{\exp(-i\cH t/\hbar)}{-j, n_{\mu} = 0}.
\end{equation}
We evaluate this to first order in the small perturbation $\cHsp$, the spin-phonon interaction. Thus,
\beq
A^{\mu}_{fi}(t)
  = \frac{1}{i\hbar} \int_0^t dt'\, \mel{j, 1}{\cHsp(t')}{-j, 0},
\eeq
where $\cHsp(t')$ is in the interaction picture generated by the rest of $\ham$, the spin-only Hamiltonian
$\mathcal{H}^{\rm s}$. Feeding in \etwo{hsp_simple}{def_qab}, and evaluating the phonon part of this matrix element, we get
\beq
A^{\mu}_{fi} = 
    - \frac{\lam_{\mu}}{\hbar} {\sum_{a,b}}^{\prime} (\kh_{a}^{\mu}e_{b}^{\mu}+\kh_{b}^{\mu}e_{a}^{\mu})
        \times \int_0^{t} dt' e^{i\om_{\mu} t'} 
           \mel{j}{ e^{-i\cHs (t-t')/\hbar} \cQ_{ab} e^{-i\cHs t'/\hbar}}{-j},
    \label{Afi_exact}
\eeq
where we have discarded the irrelevant overall phase factor $e^{-i\om_{\mu} t}$. The prime on the sum indicates the $a=b =z$ term
is to be excluded.

The remaining matrix element in \eno{Afi_exact} refers only to the spin. If we express it terms of the spectral representation of
$\cHs$, it is an excellent physical approximation to restrict ourselves to the lowest two states. It is then evident that the
matrix element is characterized by two very different energy scales, which pertain to very different physical aspects of the problem.
The two energy scales are the difference in the energy eigenvalues, and the tunneling amplitude or
matrix element. The former can be well approximated by $\veps$ when |$\veps| \gg \Dta$, a condition which is very easily satisfied. The
amplitude for tunneling is not significantly influenced by the bias, and so we may take it as its value at zero bias, $\Dta$. It is further
evident that the states with $m=-j$ and $m=j$ have near unit overlap with the higher and lower energy states respectively. Therefore,
we may write
\beq
\mel{j}{ e^{-i\cHs (t-t')/\hbar} \cQ_{ab} e^{-i\cHs t'/\hbar}}{-j}
  \simeq e^{-i\veps(t-t')/\hbar} T_{ab}(t, t')
      \label{extract_phase}
\eeq
where,
\beq
T_{ab}(t, t') = \mel{j}{ e^{-i\cHs_0 (t-t')/\hbar} \cQ_{ab} e^{-i\cHs_0 t'/\hbar}}{-j},
\eeq
with $\cHs_0$ being purely the molecular anisotropy Hamiltonian, i.e., the part of $\cHs$ without the Zeeman energy.
It follows that
\beq
A^{\mu}_{fi} \simeq 
    -\frac{\lam_{\mu}}{\hbar} 
      {\sum_{a,b}}^{\prime} (\kh_{a}^{\mu}e_{b}^{\mu}+\kh_{b}^{\mu}e_{a}^{\mu})
        \int_0^{t} dt' e^{i(\om_{\mu} + \veps/\hbar) t'} 
           T_{ab}(t,t'),
    \label{Afi_approx}
\eeq
where we have discarded another irrelevant overall phase factor, $e^{-i\veps t/\hbar}$.

The matrix element that remains in \eno{Afi_approx}, $T_{ab}$, may be said to pertain solely to the tunneling aspects of the problem,
and we expect it to be proportional to $\Dta$ (or $W$ more generally; see \eno{def_W}). This matrix element may also be written as a
path integral:
\beq
T_{ab}(t_f,t')
   =\mathcal{N}\int\mathcal{D}z\, e^{iS^{\mathrm{spin}}[z]/\hbar} 
      \mel{z(t')}{\cQ_{ab}}{{\bar z}(t')}.
   \label{Tab_pi_form}
\eeq
Here again, the sum is over all paths that go from ${\bar z}_i  = {\bar z}(t_i) = 0$ to $z_f = z(t_f) = \infty$. All other symbols
are as in \eno{ufi}, with $t_i = 0$, $t_f = t$, in particular.

As we will see below, we shall require $A^{\mu}_{fi}(t)$ for $t \gg \hbar/\veps$. If this condition is met, then,
since $\veps \ll \hbar\Om$, it is certainly true that $t \gg \Om^{-1}$. If we make the analytic continuation to imaginary times, we
will have $\Om^{-1} \ll \tau_f - \tau_i$, and we may again use our infinite-dimensional steepest descent approximation. To better
understand this, let us recall how one uses steepest descents to evaluate a one-dimensional integral such as
\beq
J = \int dx\, g(x) e^{-a f(x)}.
\eeq
If the saddle point of $f(x)$ is located at $x_0$, one expands $f(x)$ and $g(x)$ about $x_0$,
and performs the resulting Gaussian integrals to get an asymptotically valid expansion in inverse powers of $a^{1/2}$. To leading
order, we have
\beq
J \apx g(x_0) e^{-af(x_0)} \int dx\, e^{-a (x-x_0)^2 f''(x_0)/2} \simeq g(x_0) \int dx\, e^{-af(x)}.
\eeq
In other words, we may treat the factor $g(x)$ in the integrand as a constant given by its value at the saddle point.
In the case of the path integral (\ref{Tab_pi_form}), the saddle point is the entire instanton {\it path\/}, $z_d(\tau)$,
${\bar z}_d(\tau)$. Hence we must evaluate $\mel{z}{\{\cJ_a, \cJ_b\}}{{\bar z}}$ for $z(\tau) = z_d(\tau)$,
${\bar z}(\tau) = {\bar z}_d(\tau)$, do the same for $e^{iS}$, and then integrate over the Gaussian fluctuations. Finally, we
must remember to integrate over the location of the instanton center, and sum over the two types of instantons, i.e., the two
windings. Carrying out all these steps, we obtain,
\beq
T_{ab}(\tau_f, \tau_i, \tau')
 = \sum_{q = \pm 1} \int_{\tau_i}^{\tau_{f}}\!\d \tau_{c}\, e^{- (I + iq \Tta/2)/\hbar} C_{ab}(q, \tau_c)
      \times \mathcal{N}' \int'\mathcal{D}\xi\, e^{-\dta^2 S/\hbar},   \label{Tab_steep}
\eeq
where
\beq
C_{ab}(q,\tau_c) = \mel{z(\tau')}{\cQ_{ab}}{{\bar z}(\tau')},
   \qquad
      \bigl(z(\tau') = z_d(\tau'; q, \tau_c) \bigr).
\eeq
It should be further noted that in evaluating the matrix element $C_{ab}$, we retain only the leading order in $j$ because the
rest of the path integral is evaluated to this order only. This means that if we write $\bs_d$ for the vector $\bs$ along the instanton
path,
\beq
C_{ab}(q,\tau_c) = j^2 s_{d,a}(\tau'-\tau_c; q) s_{d,b}(\tau' - \tau_c; q).
\eeq

We now recall that $\tau_f > \tau' > \tau_i$. Unless $\tau'$ is very close to the end points, the $\tau_c$ integral may be evaluated
by pushing its limits to $\pm\infty$, and when that is done, the result is independent of $\tau'$. As opposed to the case of $U_{fi}$,
where the $\tau_c$ integral gave us a factor of $\tau_f - \tau_i$, here we will obtain just a number for $T_{ab}$, essentially
independent of all time arguments. Using \eno{gauss_fluct} we may write,
\beq
T_{ab} = \frac{W}{4\hbar} \sum_{q = \pm 1} e^{iq\Tta/2} \inmipi d\tau_c\, C_{ab}(q, \tau_c). \label{Tab_answer}
\eeq
It is evident that, as it should be, the tensor $T_{ab}$ is symmetric:
\beq
T_{ba} = T_{ab}.
\eeq

Let us comment briefly on the effect of the multi-instanton paths, which we have neglected above. Qualitatively, we would expect
these to modify the $e^{-i\veps(t-t')}$ phase factor that we extracted in \eno{extract_phase} by taking into account the effects
of tunneling. The primary effect would be to modulate the phase factor by an expression nearly equal to unity and varying at
frequencies of order $\Dta/\hbar$.
Within the scope of a Fermi golden rule calculation such a refinement is academic, and we are well justified in ignoring it.

As an example, let us evaluate \eno{Tab_answer} for $T_{ab}$ for the specific case of the $C = 0$, $H_x = 0$ instantons
(\ref{inst_fe8}). The integrals involved are elementary, and to leading order in $j$, we get
\beq
T_{ab} = \frac{j^2\Dta}{2\hbar\Om} \times
   \begin{cases}
       -(D - E)/E, & a\,b = x\,x, \cr
       i(D^2 - E^2)^{1/2}/E, & a\,b = x\,y, \cr
       (D + E)/E, & a\,b = y\,y. \cr
   \end{cases}
      \label{Tab_00}
\eeq
The components $T_{xz}$ and $T_{yz}$ vanish. More generally, they are odd functions of $H_x$, while the three components given above are
even functions. The component $T_{zz}$ is immaterial since the $\cQ_{zz}$ term is absent in $\cHsp$. We expect all components of this
tensor to be of order $\Dta/\hbar\Om$ as a rule just by dimensional arguments.

Returning to our transition amplitude, we have,
\beq
A^{\mu}_{fi}(t)
  = -\frac{\lam_{\mu}}{\hbar}
        \int_0^{t} dt' \, e^{i(\om_{\mu} + \veps/\hbar) t'} \,
           {\sum_{a,b}}^{\prime} (\kh_{a}^{\mu} e_{b}^{\mu} + \kh_{b}^{\mu}e_{a}^{\mu})\, T_{ab}.
\eeq
The sum is independent of $t'$ and $t$. So, for $t \gg \hbar/\veps$, the usual argument for Fermi's golden rule yields
the transition probability as
\beq
\mathcal{P}_{-j\to j}^{\mathrm{em}}(t) 
  = \sum_{\mu} |A^{\mu}_{fi}(t)|^2
  \apx \Gam_{-j \to j}^{\rm{em}} t,
\eeq
with
\beq
\Gam_{-j \to j}^{\rm em}
  = \frac{2\pi}{\hbar} \sum_{\mu} \dta(\hbar\om_{\mu} + \veps)\, 
           |\lam_{\mu}|^2 \Bigl| {\sum_{a,b}}^{\prime} (\kh_{a}^{\mu} e_{b}^{\mu} + \kh_{b}^{\mu}e_{a}^{\mu})\, T_{ab} \Bigr|^2.
\eeq

Once again, when we sum over phonon modes, we will end up integrating over all directions of $\bk$, and
summing over polarizations. The details of how to perform these operations are also in Appendix \ref{pol_avg}, and the
final result can be cast into the standard form (\ref{Gam_std_form}),
\beq
\Gam_{-j \to j}^{\rm em}
  = \frac{\Lam^2}{2\pi\rho \hbar^4} \frac{|\veps|^3}{1 - e^{-\be|\veps|}}
     \left( \frac{G^{\rm sc}_L}{15 c_L^5} + \frac{G^{\rm sc}_T}{10 c_T^5} \right),
            \label{eq:semiclassical-emissive}
\eeq
with
\begin{equation}
\begin{aligned}
G^{\rm sc}_L
 & = 4\,\left(3 |T_{xx}|^2 + 3 |T_{yy}|^2 + T_{xx} {\bar T}_{yy} + T_{yy} {\bar T}_{xx}
                            + 4 |T_{xy}|^2  + 4 |T_{xz}|^2 +4 |T_{yz}|^2 \right),\\
G^{\rm sc}_T
 & =\frac{8}{3}\,\left(2 |T_{xx}|^2 + 2 |T_{yy}|^2 - T_{xx} {\bar T}_{yy} - T_{yy}\bar{T}_{xx}
                            + 6 |T_{xy}|^2 + 6 |T_{xz}|^2 + 6 |T_{yz}|^2 \right).
\end{aligned}
    \label{GL_GT}
\end{equation}
We have restored the Bose factor, and the superscript `sc' stands for `semiclassical'.
In contrast to what we found from perturbation theory, we no longer expect $G_L = G_T$.

In general, for $C \ne 0$, $H_x \ne 0$, the spin tunneling cofactors $G^{\rm sc}_L$ and $G^{\rm sc}_T$ must be found
numerically. We can gain some insight into these quantities by considering the special case $C = H_x = 0$, when the
calculation can be done completely in closed form. It follows from \eno{Tab_00} that,
\bea
G^{\rm sc}_L &=& 8\left(\frac{j^2 \Dta}{\hbar\Om} \right)^2 \Bigl(\frac{D^2}{E^2} + \frac{1}{2}\Bigr), \label{GL_sc}\\
G^{\rm sc}_T &=& 8\left(\frac{j^2 \Dta}{\hbar\Om} \right)^2 \Bigl(\frac{D^2}{E^2} - \frac{1}{3}\Bigr). \label{GT_sc}
    \label{G_sc_C=0}
\eea
As expected, $G_L \ne G_T$. We can understand the $8j^4 \Dta^2/\hbar^2 \Om^2$ factor in a physical way, as arising from
the time scale of the instanton and the form of the spin-phonon coupling. The $j^4$ dependence is especially notable. The remaining
factor in the last parentheses may be regarded as an instanton shape factor. This shape factor is of order $10^1$ for $C=0$,
but we shall see that that is not so when $C \ne 0$. It is not easy to anticipate this fact, but it is perhaps not so
surprising in view of the large changes in the tunneling spectrum that turning on the $C$ term
produces~\cite{garg-kececioglu-jump}. In particular, note that when $C$ is turned on, $\Dta$ for $H_x = 0$ itself goes
up by about $\sim 10^3$.

To be relevant to the actual experimental situation in \Fe8, we must allow $C \ne 0$. The most important case is when
$H_x = 0$, and then we can evaluate $T_{ab}$ and thence $G_L$ and $G_T$, without excessive additional effort. We explain
how to do this in Appendix \ref{Tab_eval}.

Finally, it is natural to ask how well the semiclassical method agrees with the perturbative one. We'll see below
(in \Sno{sec:Results}) that the agreement is quite good, giving us confidence that both approaches are quantitatively
reliable.

\section{Application to $\mathrm{Fe}_{8}$}
\label{sec:Results}

We now apply our calculations specifically to $\mathrm{Fe}_{8}$, for
which $j=10$, $D=0.292\,\mathrm{K}$, $E/D=0.158$, and $C/D=-9.93\times10^{-5}$
\cite{wernsdorfer-sessoli}. For the spin-phonon interaction, we take
$\Lambda=0.25\,\mathrm{K}$ as an estimate for the coupling, and $\rho=1.92\,\mathrm{g}/\mathrm{cm}^{3}$~\cite{wieghardt84},
which with the unit cell volume of 1496$\,$\AA$^3$~\cite{wieghardt84}, and Debye temperature of $33\,$K reported in Ref.~\cite{gomes08},
implies an average sound speed of ${\bar c} = 1.4\times 10^5\,$cm/s, where the average is actually the cube root of the harmonic
mean, over all orientations and polarizations, of the cube of the phonon velocity as $k \to 0$. Within our model of an isotropic
solid, assuming $c_L/c_T = 2$ (this is a typical ratio for common organic materials --- 2.1 for polystyrene, 3.6 for polyethylene),
these numbers yield,
\begin{equation}
\begin{aligned}
\frac{\Lambda^2}{2\pi \rho \hbar^4}\times \frac{1}{15 c_L^5} &= 0.015\, \mathrm{s}^{-1} \mathrm{K}^{-3}, & & &\frac{\Lambda^2}{2\pi \rho
\hbar^4}\times \frac{1}{10 c_T^5} &= 0.72\, \mathrm{s}^{-1} \mathrm{K}^{-3}.
\end{aligned}
\end{equation}
Since the sound speeds of the two modes occur to the third (for specific heat) or fifth (for the rate) power in the denominator,
and since $c_L$ is usually larger than $c_T$, the transverse term will generally dominate for both the specific heat and the
tunneling rate, and one could neglect the longitudinal term with little change in the numbers.

Let us first compare the semiclassical and perturbative results for $\Gam^{\rm em}_{-j \to j}$.
The physical prefactors and Bose occupation factors are the same in the two methods, so
it suffices to compare the spin tunneling cofactors $G_L$ and $G_T$, i.e. $G_L^{\mathrm{pt}}$ with
$G_L^{\mathrm{sc}}$, and $G_T^{\mathrm{pt}}$ with $G_T^{\mathrm{sc}}$, noting that
$G_L^{\mathrm{pt}}=G_T^{\mathrm{pt}} \equiv G^{\rm pt}$. This is done in Table~\ref{tab:compare-noC4} for $C=0$,
and in Table~ \ref{tab:compare-C4} for $C/D = -9.93 \times 10^{-5}$. We have put $\bH = 0$, and given answers for some other
values of $j$ beside 10, keeping $E$, $D$, and $C$ the same. For the semiclassical case, we have taken $\Dta$ from a direct numerical
diagonalization of the pure spin Hamiltonian. We could in principle calculate $\Dta$ also from the instanton method, but that is not
the point of this paper.

% % % % %
\begin{table}
\begin{centering}
\begin{tabular}{|c|c|c|c|}
\hline
$j$ & $G^{\mathrm{pt}}$ & $G_L^{\mathrm{sc}}$ & $G_T^{\mathrm{sc}}$ \\[1.3ex]
\hline
\hline
$5$ & $4.479\times10^{-5}$      & $4.768\times10^{-5}$  & $4.67\times10^{-5}$\tabularnewline
\hline
$6$ & $6.969\times10^{-7}$      & $7.522\times10^{-7}$  & $7.368\times10^{-7}$\tabularnewline
\hline
$7$ & $9.397\times10^{-9}$      & $1.028\times10^{-9}$  & $1.007\times10^{-9}$\tabularnewline
\hline
$8$ & $1.141\times10^{-10}$     & $1.265\times10^{-10}$ & $1.239\times10^{-10}$\tabularnewline
\hline
$9$ & $1.280\times10^{-12}$     & $1.437\times10^{-12}$ & $1.408\times10^{-12}$\tabularnewline
\hline
$\bf{10}$ & $\bf{1.348\times10^{-14}}$ & $\bf{1.533\times10^{-14}}$& $\bf{1.502\times10^{-14}}$\tabularnewline
\hline
$11$ & $1.349\times10^{-16}$    & $1.555\times10^{-16}$ & $1.523\times10^{-16}$\tabularnewline
\hline
$12$ & $1.295\times10^{-18}$    & $1.511\times10^{-18}$ & $1.481\times10^{-18}$\tabularnewline
\hline
\end{tabular}
\par\end{centering}

\caption{Comparison between the perturbative and semiclassical answers for the spin tunneling cofactors $G_L$ and $G_T$
for $\bH = 0$ and $C/D = 0$. Note that $G^{\rm pt} = G^{\rm pt}_L = G^{\rm pt}_T$.}
\label{tab:compare-noC4}
\end{table}
% % % % %

The first point of agreement between the perturbative and semiclassical approaches is that whereas in the former,
$G_L$ and $G_T$ are strictly equal, in the latter they are not. The differences are quite small, however,
so it would not be too incorrect to speak of a single $G^{\rm sc}$ in this case too.
Second, the absolute agreement between the two approaches is quite good, with the differences being by about 10\%.
This is mainly because the perturbative method underestimates $\Dta$, and if we had used this method to provide the
input value of $\Dta$ in \eno{G_sc_C=0} and its $C\ne  0$ analog, the differences would only be about 1.5\%.

% % % % %
\begin{table}
\begin{centering}
\begin{tabular}{|c|c|c|c|}
\hline
$j$ & $G^{\mathrm{pt}}$ & $G_L^{\mathrm{sc}}$ & $G_T^{\mathrm{sc}}$ \\[1.3ex]
\hline
\hline
$5$ & $1.485\times10^{-4}$      & $1.604\times10^{-4}$  & $1.558\times10^{-4}$\tabularnewline
\hline
$6$ & $5.655\times10^{-6}$      & $6.272\times10^{-6}$  & $6.068\times10^{-6}$\tabularnewline
\hline
$7$ & $2.375\times10^{-7}$      & $2.703\times10^{-7}$  & $2.606\times10^{-7}$\tabularnewline
\hline
$8$ & $1.126\times10^{-8}$      & $1.317\times10^{-8}$  & $1.265\times10^{-8}$\tabularnewline
\hline
$9$ & $6.085\times10^{-10}$     & $7.322\times10^{-10}$ & $7.001\times10^{-10}$\tabularnewline
\hline
$\bf{10}$ & $\bf{3.755\times10^{-11}}$ & $\bf{4.655\times10^{-11}}$& $\bf{4.431\times10^{-11}}$\tabularnewline
\hline
$11$ & $2.641\times10^{-12}$    & $3.377\times10^{-12}$ & $3.200\times10^{-12}$\tabularnewline
\hline
$12$ & $2.109\times10^{-13}$    & $2.788\times10^{-13}$ & $2.628\times10^{-13}$\tabularnewline
\hline
\end{tabular}
\par\end{centering}

\caption{Same as Table~\ref{tab:compare-noC4}, except with $C/D = -9.93\times10^{-5}$.}
\label{tab:compare-C4}
\end{table}
% % % % %

The second key point is that for $j = 10$, turning on the $C$ term increases $\Dta$ by about $10^3$. Since $G$ increases
by about the same factor, and it is proportional to $\Dta^2$, this means that the shape factor decreases by $\sim10^{-3}$.

Next, let us consider how the rate varies with $H_z$ (or $\veps$). It isuseful to introduce the reduced magnetic field,
\beq
\bh = \frac{g\mu_B \bH}{Dj}.
\eeq
The dominant dependence is $h_z^3$ from the phase space plus Bose
factor of $|\veps|^3/(1 - e^{-\be|\veps|})$. There may, however, be an additional dependence in the $G$'s. Finding this
dependence with the instanton method is an unsolved problem (since the classical energy minima are no longer degenerate), but it is
straightforward in the perturbative
approach, where it arises from the $\veps$ dependence of the energy denominators. We show this dependence in 
Fig.~\ref{fig:phi-compare} for $j=10$ with $C/D = -9.93\times10^{-5}$, and $h_x=0$. We see that there is indeed a
weak dependence on
$\varepsilon$ at small fields. The change from the $h_z = 0$ value rises from less than 3\% at $h_z = 0.02$ to 18\% at
$h_z = 0.06$. Note that we hit the next resonance (between $m=-10$ and $m=9$) at $h_z \simeq 0.1$. We also show
the semiclassical answers for $G_L$ and $G_T$ at $h_z = 0$; it can be seen that they serve as reasonable estimates for the entire range of
$h_z$ values shown.

% % % % %
\begin{figure}
\begin{centering}
\includegraphics[scale=0.75]{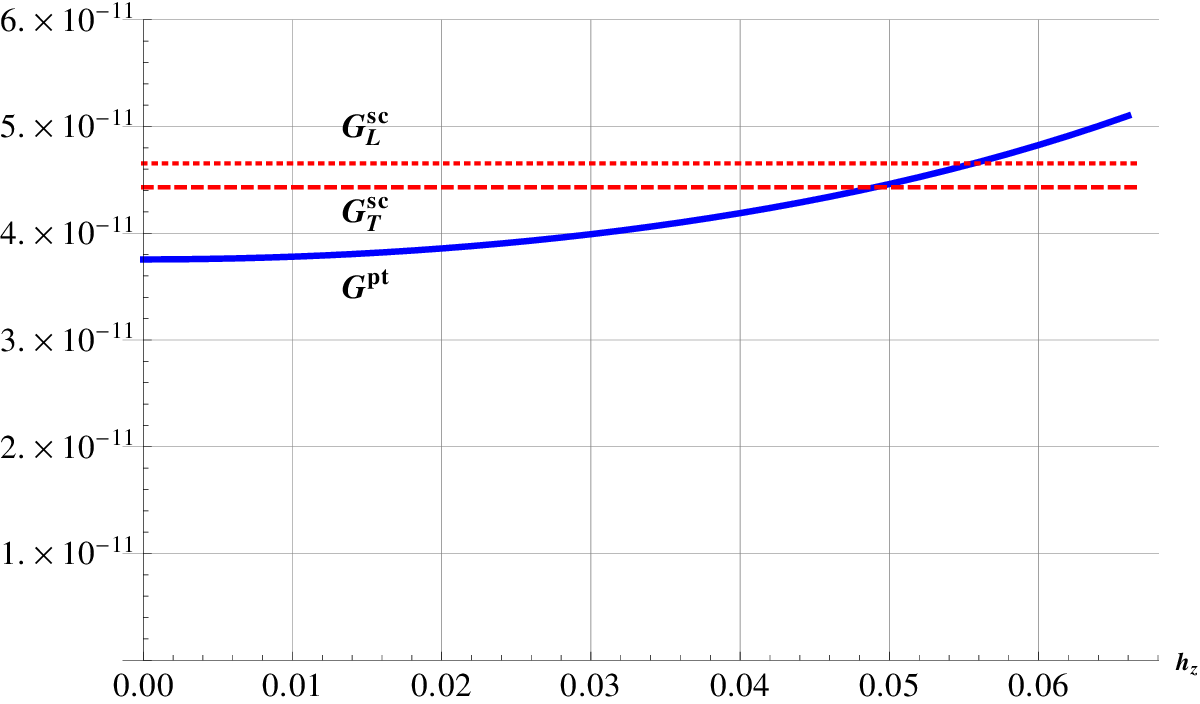}
\par\end{centering}
\caption{Plot of $G^{\mathrm{pt}}$ as a function of external applied field $h_z$ along the easy axis (solid blue line),
for $j=10$ with $C/D=-9.93\times10^{-5}$, and $h_x = 0$.  
Note that the resonance between $m=-j$ and $m=j-1$ when $j=10$ is located at $h_z = 0.1$. We also show the
$h_z = 0$ values of
$G_L^{\mathrm{sc}}$ (dotted red line) and $G_T^{\mathrm{sc}}$ (dashed red line), with all other parameters the same.
\label{fig:phi-compare}}
\end{figure}
% % % % %

Lastly, we consider the general case when both $h_x$ and $h_z$ are non-zero. In  \fno{fig:phi-compare-2} we look at
plots of $G^{\mathrm{pt}}$ vs $h_z$, for several values of $h_x$. (The curve for $h_x = 0$ is the same as in \fno{fig:phi-compare}.) Note that with increasing $h_x$, the rate becomes
smaller as a whole, and its $h_z$ dependence
becomes more pronounced. The overall drop is because of interference. At $h_z = 0$, in particular, we will
find that with increasing $h_x$, the $G$'s vanish, and so, consequently, does $\Gamma$. This is
because we are approaching a \emph{diabolical point} in $\boldsymbol{H}$ space, where the pure spin problem
tunnel splitting $\Delta$ vanishes. (For
$\mathrm{Fe}_8$, the first such point along $h_z=0$ is at $h_x=0.09345$, or $H_x = 0.203 \mathrm{T}$.)

% % % % %
\begin{figure}
\begin{centering}
\includegraphics[scale=0.75]{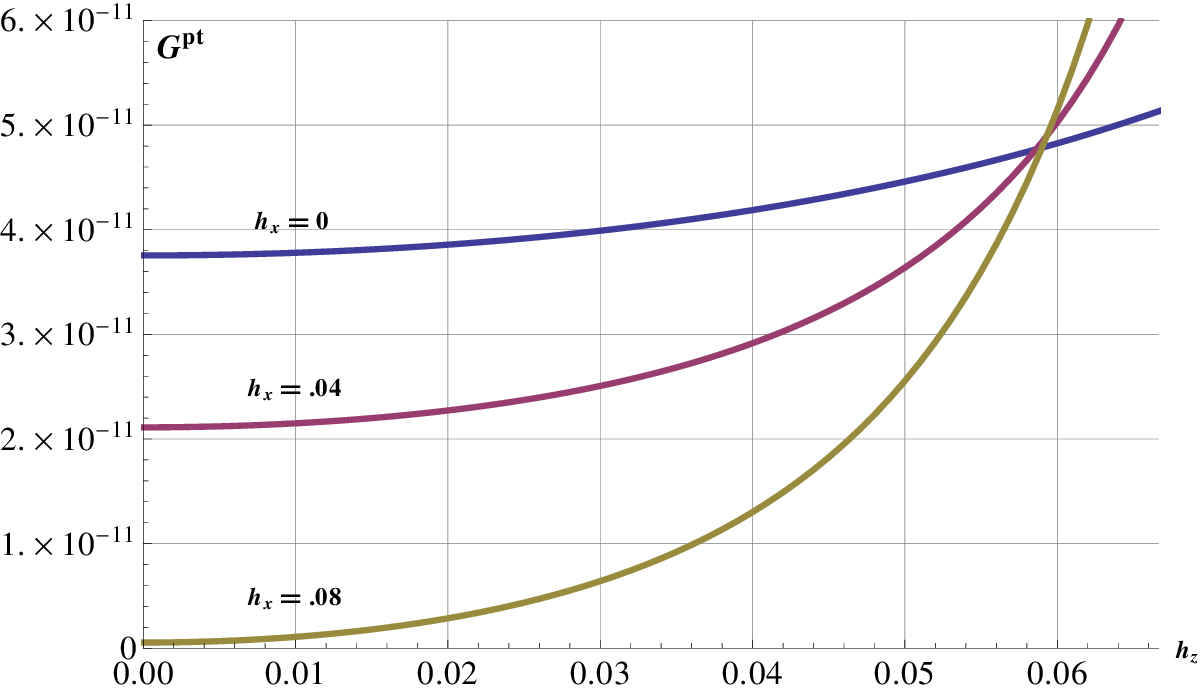}
\par\end{centering}

\caption{Plot of $G^{\mathrm{pt}}$ as a function of external applied field $h_z$, for various values of $h_x$, and all other
parameters at the values for \Fe8. Note that the intersections exhibited around $h_z = 0.06$ do not
all occur at the same location. \label{fig:phi-compare-2}}
\end{figure}
% % % % %

\section{Conclusions}
\label{concs}

We have calculated the phonoemissive and phonoabsorptive tunnelling rates in single molecule magnets by two methods, perturbation
theory, and instantons. In addition, we have offered interpretations of the calculations in terms of the spin tunneling cofactor,
and the instanton shape factor, and attempted to isolate the effects of the various ingredients that go into the rates.
The two methods agree quite well with each other. 

How well do our calculations compare with previous
work~\cite{politi95,garg-adp}?  Reference~\cite{politi95} is concerned entirely with the case where $E = 0$, $C \ne 0$, and its
general approach is as follows. This paper employs not the spin states $\ket{j,m}$, which are the eigenstates of
$\mathcal{J}_z$, but rather the states $\ket{j,m^*}$, which are linear combinations of the exact eigenstates of the pure spin
Hamiltonian such that the wavefunction is localized in one well or the other. In most cases of interest one works with the state
$\ket{j,\pm j^*}$, which are formed from the two lowest lying eigenstates. These states were determined approximately by considering
just the exponential fall-off into the opposite well, ignoring their modification due to the actual shape of the potential.
We believe that the method would also give a good description of the physics for the case $E \ne 0$, and $C = 0$.
When $E$ and $C$ are both nonzero however, as is the case for Fe$_8$, and when both terms are important, then this method is
difficult to apply, and in light of our finding that the instanton shape factor is not a simple number, it is unclear that assuming
a simple exponential drop-off in the wavefunctions would be adequate. If additional anistropy due to $H_x$ is added, the technique
is even harder to apply.

We can also try to understand the matrix elements between states $\ket{j,\pm j^*}$ in terms of the semiclassical
instanton approach. Since the instantons are the classical trajectories for the full spin Hamiltonian
$\mathcal{H}^{\mathrm{s}}$ and not just its diagonal part $\mathcal{H}_0 - \mathcal{H}^{\mathrm{p}}$, the first order perturbation in
$\cHsp$ via these instantons captures the intentions of $\ket{j,\pm j^*}$. Again, however, it is unclear if one could capture
situations with nontrivial instanton shape factors in this way.

In Ref.~\cite{garg-adp} on the other hand, $\Gam$ was obtained by doing a second order Fermi golden rule calculation taking
only the $m=\pm 9$ states as intermediates, and using experimental data~\cite{wernsdorfer-sessoli} to obtain the amplitudes
$\Dta_{-10,9}$ and $\Dta_{9,-10}$. This is like the perturbative calculation of the present paper, except that $\cHsp$ is restricted
to act at either the first or last step of the avaliable paths. We do not have a strong justification for this approximation,
even though our present answers are broadly consistent with it. Still, it relies on the availability of of the $\Dta_{-10,9}$
and $\Dta_{9, -10}$ amplitudes from experimental data in order to finesse large parts of the complete cacluation, and in that
regard it is not a complete theory.

Lastly let us consider the absolute value of the tunneling rate. Here we return to $H_x=0$. The most significant term in
\eno{Gam_pert} is the transverse spin tunneling cofactor, $G_T^{\mathrm{pt}}$, which is independent of the unknowns
$\Lambda$, $c_L$, $c_T$, and $\rho$. Over the relevant values of $H_z$, $G_T^{\mathrm{pt}}$ is of order
$10^{-11}$. If we take $h_z$ to be such that the local bias is $\varepsilon \sim 0.1\mathrm{K}$, then at zero temperature the tunneling rate
is $~2.7\times 10^{-14}\,\mathrm{s}^{-1}$. As argued in Ref.~\cite{garg-adp}, even with uncertainties in the magnetoelastic coupling
and sound speeds, its order of magnitude is too small to explain the experimental
magnetization data, and so the problem of how single molecule magnets get magnetized starting from a demagnetized state remains open.
It also remains an open question if the rates calculated in this paper could be measured directly.

\acknowledgments
This work was supported in part by the NSF via grant number PHY-0855323 AR at Urbana, and also
in part by the NSF via grant number PHY-0854896 at Evanston.
We are indebted to Michael Stone for many useful discussions. 

\appendix
\section{Summing over Phonon Polarizations and Directions}
\label{pol_avg}

The purpose of this Appendix is to show how to sum over phonon polarizations and directions.
Throughout the Appendix, we omit the label $\bk$ in the polarization vectors, and write just $e^s_-$ and $e^s_z$ instead
of $e^{\bk s}_-$ and $e^{\bk s}_z$. This improves the readability of the formulas at little cost since it is easy to
remember that we are dealing with a mode with wavevector $\bk$. We shall restore the $\bk$ label when necessary or useful.

\subsection{Basic identities}
\label{basic_ids}

The fundamental fact on which we rely is that for any $\bk$, the polarization vectors are a
complete set. That is, with $a$ and $b$ being Cartesian indices,
\beq
\sum_s e^{\bk s}_a e^{\bk s}_b = \dta_{ab}.
\eeq
The sum over the polarization label, $s$, can be taken to run over the longitudinal mode ($s = L$), and the two transverse
modes ($s= T_1, T_2$ or $s \ne L$). For $s =L$, ${\bf e}^{\bk s} = \khat$. Hence,
\beq
e^L_a e^L_b = \kh_a \kh_b. \label{elel}
\eeq
It follows that
\beq
\sum_{s \ne L} e^s_a e^s_b
   = \dta_{ab} - \kh_a \kh_b.      \label{etet}
\eeq
Secondly,
\beq
\int d^2\hat\bk\, \kh_z^2 = \frac{4\pi}{3}, \quad
\int d^2\hat\bk\, \kh_z^4 = \frac{4\pi}{5},
     \label{avg_kz_powers}
\eeq
which are special cases of
\beq
\int d^2\hat\bk\, \kh_a \kh_b = \frac{4\pi}{3} \dta_{ab}, \quad
\int d^2\hat\bk\, \kh_a \kh_b \kh_c \kh_d
     = \frac{4\pi}{15} (\dta_{ab} \dta_{cd} + \dta_{ac}\dta_{bd} + \dta_{ad}\dta_{bc}).
     \label{avg_k_tensors}
\eeq
Equations~(\ref{elel})--(\ref{avg_k_tensors})  are our basic working tools.

\subsection{Sums for pertubative calculation}
\label{perturb_case}

For the perturbative calculation, we must carry out the sum over $\khat$ and $s$ for $|\cF^{\bk s}|^2$, where
$\cF^{\bk s}$ is given by \eno{F_structure}, which we reproduce here for easy reference:
\beq
\cF^{\bk s} =-i \lam_{\bk s} \bigl[ \kh_- e^s_- M_2 + (\kh_- e^s_z + \kh_z e^s_-) M_1 \bigr].
\eeq
We have
\beq
\frac{|\cF^{\bk s}|^2}{|\lam_{\bk s}|^2}
  = |\kh_- e^s_-|^2 |M_2|^2 + |\kh_- e^s_z + \kh_z e^s_-|^2 |M_1|^2
        + \bigl((\kh_- e^s_-)(\kh_+ e^s_z + \kh_z e^s_+) M_2 {\bar M}_1 + {\rm c.c.}\bigr),
\eeq
where the overbar denotes the complex conjugate.
We consider the coefficients of $|M_2|^2$, $|M_1|^2$, and $M_2 {\bar M}_1$ separately.
The quantities $|M_2|^2$, $|M_1|^2$, and $M_2 {\bar M}_1$ themselves depend on the phonon mode only 
through $|\bk|$ and whether $s=L$ or $s \ne L$. Thus, for both the longitudinal and nonlongitudinal cases,
it is possible to integrate over the wavevector orientation $\khat$, obtaining coefficients that are pure
numbers.

{\bf Coefficients of} $|M_2|^2$: We have
\beq
|\kh_- e^s_-|^2 = (\kh_x^2 + \kh_y^2) (e^s_x e^s_x + e^s_y e^s_y).
\eeq
Hence,
\bea
|\kh_- e^L_-|^2 
    &=& 1 - 2 \kh_z^2 + \kh_z^4, \\
\sum_{s \ne L} |\kh_- e^s_-|^2 
    &=& 1 -  \kh_z^4,
\eea
and
\bea
\int d^2\hat\bk\, |\kh_- e^L_-|^2 
    &=& \frac{32\pi}{15}, \label{avg_1}\\
\int d^2\hat\bk\, \sum_{s \ne L} |\kh_- e^s_-|^2 
    &=& \frac{48\pi}{15} \label{avg_2}.
\eea

{\bf Coefficients of} $|M_1|^2$: Now,
\beq
|\kh_- e^s_z + \kh_z e^s_-|^2
   = (\kh_x^2 + \kh_y^2)e^s_z e^e_z + \kh_z^2  (e^s_x e^s_x + e^s_y e^s_y) + \kh_- \kh_z e^s_z e^s_+ + \kh_+ \kh_z e^s_z e^s_-.
\eeq
A short and straightforward calculation shows that,
\bea
|\kh_- e^L_z + \kh_z e^L_-|^2
    &=& \kh_z^2 - \kh_z^4, \\
\sum_{s \ne L} |\kh_- e^s_z + \kh_z e^s_-|^2
    &=& 1 - 3\kh_z^2 + 4\kh_z^4.
\eea
Hence,
\bea
\int d^2\hat\bk\, |\kh_- e^L_z + \kh_z e^L_-|^2
    &=& \frac{32\pi}{15}, \label{avg_3}\\
\int d^2\hat\bk\, \sum_{s \ne L} |\kh_- e^s_z + \kh_z e^s_-|^2
    &=& \frac{48\pi}{15}. \label{avg_4}
\eea

{\bf Coefficients of} $M_2 {\bar M}_1$: This time, we need
\beq
(\kh_- e^s_-)(\kh_+ e^s_z + \kh_z e^s_+) = (\kh_x^2 + \kh_y^2) e^s_- e^s_z + \kh_- \kh_z (e^s_x e^s_x + e^s_y e^s_y).
\eeq
Therefore,
\bea
(\kh_- e^L_-)(\kh_+ e^L_z + \kh_z e^L_+)
    &=&  \kh_z \kh_- (1 - \kh_z^2), \\
\sum_{s \ne L} (\kh_- e^s_-)(\kh_+ e^s_z + \kh_z e^s_+) 
    &=&  \kh^3_z \kh_-,
\eea
but both these terms vanish upon integrating over $\hat\bk$. Hence there is no term involving $M_2 {\bar M}_1$, or its complex conjugate,
${\bar M}_2 M_1$.

Equations (\ref{avg_1}), (\ref{avg_2}), (\ref{avg_3}), and (\ref{avg_4}) lead immediately to \eno{FL_FT}.

\subsection{Sums for semiclassical calculation}
\label{semiclassical_case}

The sums over $s$ and $\khat$ required in \Sno{sec:Semiclassical-Calculation} are of a slightly different form,
and best performed as follows.

For phonon polarization $s$, we define,
\begin{equation}
g_{s}=\sum_{a,b}(\hat{k}_{a}e_{b}^{s}+\hat{k}_{b}e_{a}^{s})T_{ab},
\end{equation}
and
\begin{equation}
\begin{aligned}
\tG_{L} & \equiv \int d^2{\khat}\, g_{L}\bar{g}_{L}, &  &  &
\tG_{T} & \equiv \int d^2{\khat}\, \sum_{s\neq L} g_{s}\bar{g}_{s}.
\end{aligned}
\end{equation}
It is not difficult to see that with these definitions, the result for the phonoemissive rate will take on
the standard form (\ref{Gam_std_form}) with the spin tunneling cofactors given by
\beq
G^{\rm sc}_L = \frac{15}{4\pi} \tG_L, \quad G^{\rm sc}_T = \frac{10}{4\pi} \tG_T.
\eeq

In the calculations that follow, we need only remember that $T_{ab}$ is independent of $\khat$, and that $T_{ab} = T_{ba}$. Let us
first consider $\tG_L$. Then, $g_L = 2 \hat{k}_a \hat{k}_b {\bar T}_{ab}$ with an implicit sum over repeated indices, and
\bea
\tG_L &=& 4\int d^2\khat \, \hat{k}_a \hat{k}_b \hat{k}_c \hat{k}_d \, T_{ab} {\bar T}_{cd} \nnu\\
      &=& \frac{16\pi}{15}\, (T_{aa} {\bar T}_{bb} + 2 T_{ab} {\bar T}_{ab}).
\eea
To obtain the last line, we used \eno{avg_k_tensors} and summed over indices c and d.

To find $\tG_T$, we use the completeness of the $e^s_a$'s to first show that
\bea
\sum_s |g_s|^2
    &=& 4 \sum_s  \hat{k}_a e^s_b T_{ab} \hat{k}_c e^s_d {\bar T}_{cd} \nnu\\
    &=& 4 \hat{k}_a T_{ab} \hat{k}_c {\bar T}_{cd} \dta_{bd} \nnu\\
    &=& 4 \hat{k}_a T_{ab} \hat{k}_c {\bar T}_{cb}.
\eea
Integrating over $\khat$, we obtain
\beq
\tG_L + \tG_T = \frac{16\pi}{3} T_{ab} {\bar T}_{ab}.
\eeq
It follows that,
\beq
\tG_T = \frac{16\pi}{15}\,(3 T_{ab}{\bar T}_{ab} - T_{aa} {\bar T}_{bb}).
\eeq
Writing out the implicit sums explicitly, and setting $T_{zz}$ to zero, we have
\begin{equation}
\begin{aligned}
\tG_{L}
 & =\frac{16\pi}{15}\,\left(3 |T_{xx}|^2 + 3 |T_{yy}|^2 + T_{xx} {\bar T}_{yy} + T_{yy} {\bar T}_{xx}
                            + 4 |T_{xy}|^2  + 4 |T_{xz}|^2 +4 |T_{yz}|^2 \right),\\
\tG_{T}
 & =\frac{16\pi}{15}\,\left(2 |T_{xx}|^2 + 2 |T_{yy}|^2 - T_{xx} {\bar T}_{yy} - T_{yy}\bar{T}_{xx}
                            + 6 |T_{xy}|^2 + 6 |T_{xz}|^2 + 6 |T_{yz}|^2 \right).
\end{aligned}
    \label{GL_GT1}
\end{equation}

This is how \etwo{eq:semiclassical-emissive}{GL_GT} are obtained.
%
% End of Appendix A
%

\section{Evaluation of $T_{ab}$ for $H_x = 0$, $C \ge 0$}
\label{Tab_eval}

In this Appendix, we discuss how to evaluate the expression (\ref{Tab_answer}) for $T_{ab}$ for the \Fe8 problem,
especially for $C\ne 0$. Instantons with $C\ne 0$ were considered in Ref.~\cite{garg-kececioglu-jump}, but the
focus there was on the non-interfering ones with $H_x \ne 0$. Here, we are more interested in the {\it time dependence\/}
of the interfering instantons when $H_x = 0$.

In the absence of any external field, the \Fe8 Hamiltonian is
\begin{equation}
\mathcal{H}
   = -D\mathcal{J}_z^{2} + E(\mathcal{J}_x^{2}-\mathcal{J}_y^{2}) + C(\mathcal{J}_{+}^{4}+\mathcal{J}_{-}^{4}).
     \label{ham_fe8_0}
\end{equation}
The equations of motion for the instanton entail $H(z,\baz) \equiv \mel{z}{\ham}{\baz}$.
Keeping only the leading order in $j$ for the coherent state expectation value of each term, we obtain
\beq
H(z,\baz) = 2j^2 D \left[ -\hf s_z^2(z,\baz) + \frac{\al}{2} (s_x^2(z,\baz) - s_y^2(z,\baz))
                        + \gam \frac{z^4 +\baz^4}{(1+z\baz)^4} + \hf \right].
\eeq
The additive constant in $H$ has been adjusted to make the minimum energy equal to zero, and we have introduced
\beq
\al = \frac{E}{D}, \quad
\gam = 8j^3 \frac{C}{D}.
\eeq

Instead of $z$, $\baz$, or $s_a$, it is best to work in Archimedean cylindrical coordinates $s_z$ and $\phi$, given by
\bea
& z = {\dst \sqrt{\frac{1 + s_z}{1 - s_z}}} e^{i\phi},
& \baz = \sqrt{\frac{1 + s_z}{1 - s_z}} e^{-i\phi}, \\
& s_x = \sqrt{1 - s_z^2} \cos\phi,
& s_y = \sqrt{1 - s_z^2} \sin\phi.
\eea
In these coordinates,
\beq
H(s_z, \phi)
  = 2j^2 D (1-s_z^2) \left[\hf + \frac{\al}{2} \cos(2\phi) + \frac{\gam}{8} (1-s_z^2) \cos(4\phi) \right],
\eeq
and the equations of motion read
\beq
\frac{ds_z}{dt} = -\frac{1}{j} \frac{\ptl H}{\ptl\phi}, \quad
\frac{d\phi}{dt} =  \frac{1}{j} \frac{\ptl H}{\ptl s_z}.
\eeq

We now continue to imaginary time via $t \to -i\tau$, and also define $\bphi = i\phi$. Then
\beq
s_x = \sqrt{1 - s_z^2} \cosh\bphi, \quad
s_y = -i \sqrt{1 - s_z^2} \sinh\bphi, 
\eeq
the Hamiltonian reads
\beq
H(s_z, \bphi)
  = 2j^2 D (1-s_z^2) \left[\hf + \frac{\al}{2} \cosh(2\bphi) + \frac{\gam}{8} (1-s_z^2) \cosh(4\bphi) \right],
     \label{ham_inst}
\eeq
and the equations of motion become
\beq
\frac{ds_z}{d\tau} = -\frac{1}{j} \frac{\ptl H}{\ptl\bphi}, \quad
\frac{d\bphi}{d\tau} =  \frac{1}{j} \frac{\ptl H}{\ptl s_z}.
\eeq

The usual program now is to exploit the conservation of energy to find the instantons. Setting $H(s_z, \bphi) = 0$ will give
us the solution(s) for $\cosh\bphi (s_z)$, which can be fed into the equation for $ds_z/d\tau$ and used to solve for $s_z(\tau)$.
It pays to consider the $C=0$ case separately first, as that helps avoid pitfalls in the $C\ne 0$ case.

{\bf Case I.} $C = 0$:
Setting $H = 0$ with $C = 0$ ($\gam =0$) in \eno{ham_inst} yields the non-tunneling solutions $s_z = \pm 1$, as well as the tunneling one:
\beq
\cosh(2\bphi) = -\frac{1}{\al}.
\eeq
Along this solution, $\al \sinh(2\bphi) = \pm \sqrt{1 - \al^2}$, and the equation of motion becomes
\beq
\frac{d s_z}{d\tau} = \pm 2j D \sqrt{1-\al^2} (s_z^2 - 1),
\eeq
which, with $\hbar\Om \equiv 2jD \sqrt{1-\al^2}$, gives
\beq
s_z = \pm \tanh\Om(\tau - \tau_c).
\eeq
The two signs here describe tunneling in opposite directions, from $-\zhat$ to $\zhat$, and from $\zhat$ to $-\zhat$, {\it not\/} the
two interfering instantons. Let us pick the one with the plus sign, wherein $s_z$ goes from $-1$ to $1$. For this solution, we
may obtain the other two Cartesian components of $\bs$ by noting that
\beq
\cosh^2\bphi = \tshf \bigl(\cosh(2\bphi) + 1 \bigr), \quad
\sinh^2\bphi = \tshf \bigl(\cosh(2\bphi) - 1 \bigr).
\eeq
The oppositely winding, interfering solutions arise from the two different branches of the square root, which introduces a relative phase of
$\pi$ between them. In particular, for $\al < 1$, both $\cosh(2\bphi) \pm 1$ will be negative. (Note that since we have chosen
$\sinh(2\bphi) = +\sqrt{1 - \al^2}/\al$, once a sign for $\cosh\bphi$ is picked, that for $\sinh\bphi$ is also fixed.) 

{\bf Case II.} $C \ne 0$:
Setting $H = 0$ with $C \ne 0$ leads to the tunneling solutions
\beq
\cosh(2\bphi)
 = -\frac{\al \pm \sqrt{\al^2 - \tshf\gam \bigl(1 - s_z^2)(4 - \gam(1 - s_z^2)\bigr)}}
         {\gam (1-s_z^2)}.
\eeq
Of these two solutions, we must choose the one with the relative ``$-$'', since in that case (see \fno{fig-cosh2phi}),
\beq
\lim_{s_z \to \pm 1} \cosh(2\bphi) = -\frac{1}{\al},
\eeq
and since this is a finite quantity, $s_x$ and $s_y$ will vanish as $s_z \to \pm 1$. We are thus assured that this solution will
approach $\bs = \pm\zhat$ at the end points, which, as shown in Ref.~\cite{garg-kececioglu-jump} is the hallmark of the {\it interfering
instantons\/}. The other solutions do not approach $\pm\zhat$, i.e., do not start and end on the real two-sphere, and are thus the
noninterfering or jump instantons.

% % % % %
\begin{figure}
\begin{centering}
\includegraphics[scale=0.70]{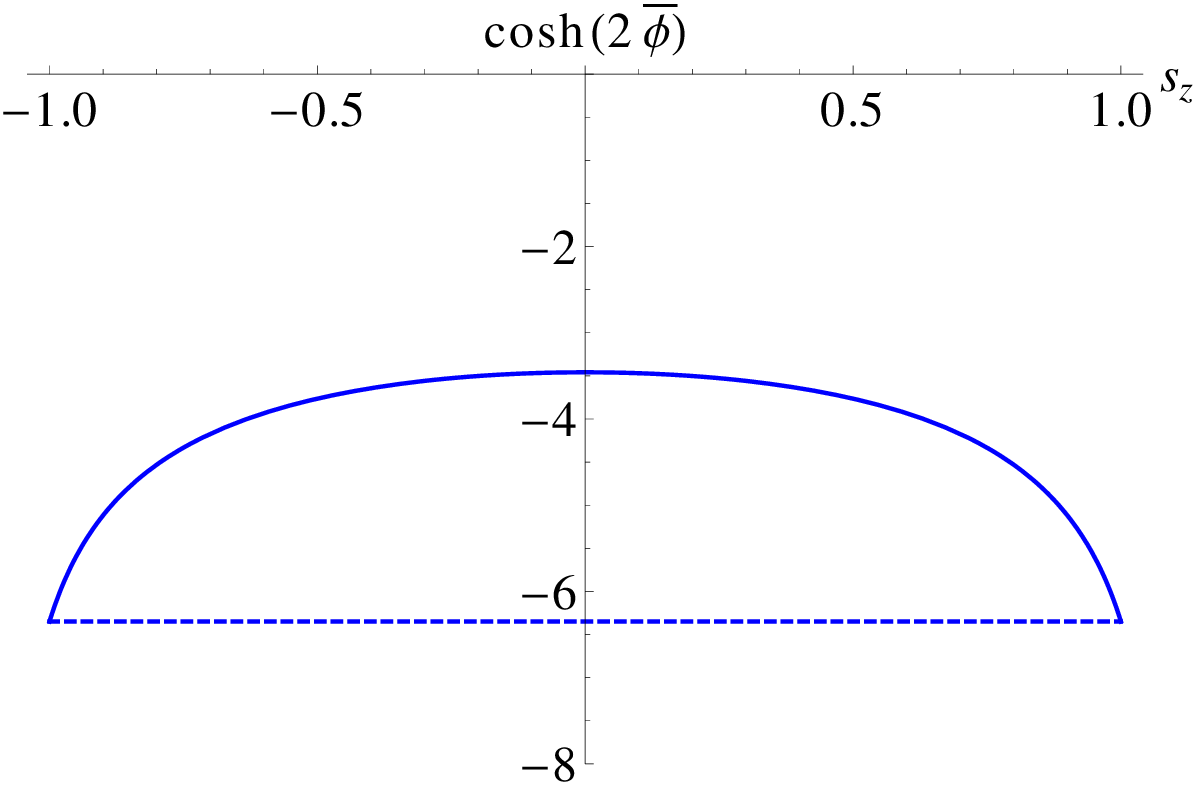}
\par\end{centering}
 
\caption{Plot of $\cosh(2\bar{\phi})$ as a function of $s_z$, for the case with $C/D=0$ (dashed lines) and $C/D = -9.93\times 10^{-5}$
(solid lines). The other parameters are those for Fe$_8$. \label{fig-cosh2phi}}
\end{figure}
% % % % % 

Another way to think about this is to take the limit $\gam \to 0$ since the term $(1-s_z^2)$ is always paired up with $\gam$. Again, the
solution with a relative ``$-$'' is the continuation of the $\gam = 0$ case, whereas the ``$+$'' solution diverges as
$\gam \to 0$, and is new. This equation reveals the singular nature of the fourth order perturbation.

From now on, we consider the interfering instantons only since these are the ones for which the real part of action is the least.
Feeding that solution for $\cosh(2\bphi)$ into the equation of motion, we obtain a differential equation of the form
\beq
\frac{ds_z}{d \tau} = f(s_z), \label{ode_sz}
\eeq
where the functional form of $f$ is very cumbersome, but (for the parameters relevant to \Fe8 at the least) has the following
properties (see \fno{fig-f-fn}):

{{\narrower
1. $f(-s_z) = f(s_z)$.

2. $f(\pm 1) = 0.$

3. $f(s_z)$ is real for $s_z \in [-1, 1]$.
}}

% % % % %
\begin{figure}
\begin{centering}
\includegraphics[scale=0.70]{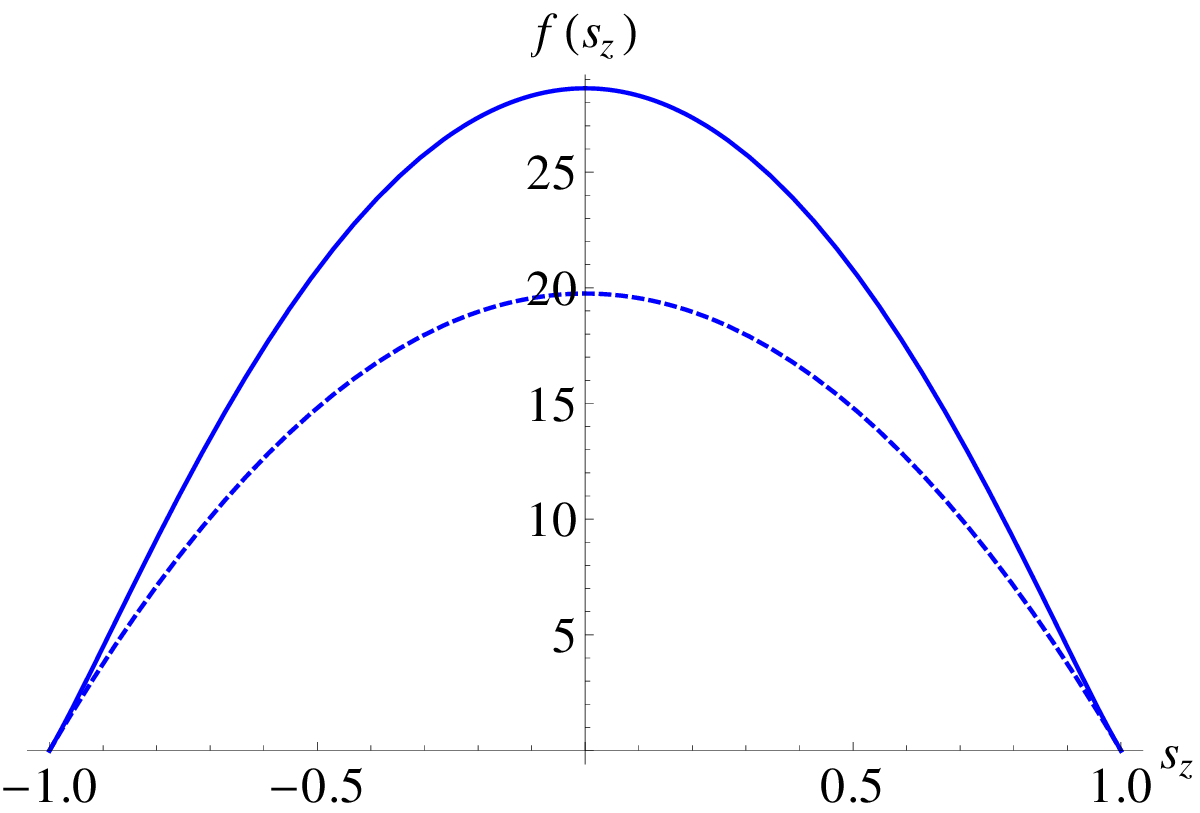}
\par\end{centering}

\caption{Plot of $f(s_z)$ in (\ref{ode_sz}) as a function of $s_z$, for the case with $C/D=0$ (dashed line) and $C/D=-9.93\times 10^{-5}$
(solid line), with $H_x=0$ in both cases. The other parameters are as for \Fe8. \label{fig-f-fn}}
\end{figure}
% % % % %

This shows that we can find a {\it real\/} solution for $s_z(\tau)$ that connects the endpoints $\pm 1$. Therefore, instead of
solving \eno{ode_sz}, we can use $ds_z/f(s_z)$ as a change of measure for $d\tau$. We use this to evaluate the integral in
\eno{Tab_answer}. We first use the fact that the instanton depends only on $\tau - \tau_c$ to write
\beq
\inmipi d\tau_c\, s_a(q, \tau_c; \tau) s_b(q, \tau_c; \tau)
 = \inmipi d\tau\, s_a(q, \tau_c; \tau) s_b(q, \tau_c; \tau).
\eeq
We then transform the integral to one over $s_z$, from $-1$ to $+1$, {\it along the real axis\/},
\beq
\int_{-1}^1 \frac{ds_z}{f(s_z)} s_a(q, \tau_c; \tau) s_b(q, \tau_c; \tau).
\eeq
Since $s_x$ and $s_y$ can be obtained as functions of $s_z$, the numerical evaluation of this integral is straightforward.

One immediate consequence of this procedure is that since $s_x$, $s_y$, and $f(s_z)$ are all even functions of $s_z$, while
$s_z$ is trivially odd, $T_{xz}$ and $T_{yz}$ vanish by oddness of the integrand. This agrees with our perturbative result
that for integer $j$ the $\Dta m = 1$ terms vanish when $H_x = 0$, and with the explicit instanton calculation for $C=0$.
Furthermore, the remaining integrals in $T_{xx}$, $T_{xy}$, and $T_{yy}$ are insensitive to the sign of $q$ (the winding), since
either $s_x$ and $s_y$ both have a positive sign, or both have a negative sign. Hence the sum over $q$ will once again produce
the tunnel splitting $\Dta$ precisely.

We conclude the Appendix by noting that our symmetry arguments fail when $H_x \ne 0$. The function $f(s_z)$ is no longer purely
real for $s_z \in [-1, 1]$, so we cannot claim that the trajectory $s_z(\tau)$ lies entirely along the real axis from $-1$ to $1$.
The $\Dta m = 1$ terms no longer vanish, and the sum over windings produces $W \cos(\tshf\Tta + \zeta)$, where $\zeta$
is a nonzero additional phase, so while the $T_{ab}$'s are still of order $\Dta$, they are not all strictly proportional to it.


\begin{thebibliography}{8}
\bibitem{Lam_form}
This is the simplest form consistent with the general symmetry properties of the magnetoelastic tensor,
$\Lam_{abcd} = \Lam_{bacd} = \Lam_{abdc}$. One might wonder if a term like $\Lam'\dta_{ab}\dta_{cd}$ should
also be included. The answer is no, because it would be proportional to $\bcJ\cdot\bcJ$ which is a constant.
Thus, such a term would involve only the phonons, and because it was linear in the strain, would imply the existence
of a nonzero strain in equilibrium. Since we wish to regard the equilibrium state as unstrained, i.e., measure strain
with reference to this state, we set $\Lam' = 0$.
\bibitem{dps}
Since the different terms can have different signs (or complex-valued phases in general), the addition of contributions
from different paths can give
rise to interference. In the perturbative point of view, it is this interference that underlies the diabolical points
in the tunneling spectrum, and as found in Ref.~\cite{park-garg-perturbative} keeping the lowest order terms of each
qualitatively different class of discrete paths gives a quantitatively good account of the locations of the diabolical
points.
\end{thebibliography}
\end{document}